\documentclass[11pt,a4paper]{article}

\usepackage{jheppub}
\usepackage{amsfonts}
\usepackage{amssymb}
\usepackage{comment}
\usepackage{epsfig}
\usepackage{graphicx}
\usepackage{amsmath}
\usepackage{tabu}

\newcommand{\bea}{\begin{eqnarray}}
\newcommand{\eea}{\end{eqnarray}}
\newcommand{\be}{\begin{equation}}
\newcommand{\ee}{\end{equation}}
\newcommand{\ba}{\begin{align}}
\newcommand{\ea}{\end{align}}
\newcommand{\comments}[1]{}

\def\SM{{\scriptscriptstyle \rm SM}}
\def\KK{{\scriptscriptstyle \rm KK}}
\def\dS{{\scriptscriptstyle \rm dS}}
\def\lp{{\scriptscriptstyle \rm loop}}

\def\cv{{{\cal{V}}}}

\newcommand\vo{{\mathcal{V}}}

\newcommand{\mc}{\mathcal}

\newcommand{\beqa}{\begin{eqnarray}}
\newcommand{\eeqa}{\end{eqnarray}}

\title{Sequestered de Sitter String Scenarios: Soft-terms}

\author[1]{Luis Aparicio,}
\author[1,2,3]{Michele Cicoli,}
\author[4]{Sven Krippendorf,}
\author[5]{Anshuman Maharana,}
\author[2,3]{Francesco Muia,}
\author[1,6]{Fernando Quevedo}

\affiliation[1]{ICTP, Strada Costiera 11, Trieste 34014, Italy}
\affiliation[2]{Dipartimento di Fisica e Astronomia, Universit\`a di Bologna, \\ via Irnerio 46, 40126 Bologna, Italy}
\affiliation[3]{INFN, Sezione di Bologna, Italy}
\affiliation[4]{Bethe Center for Theoretical Physics and Physikalisches Institut der \\ Universit\"at Bonn, Nussallee 12, 53115 Bonn, Germany}
\affiliation[5]{Harish Chandra Research Institute, Chhatnag Road, Jhunsi, Allahabad, UP 211019, India}
\affiliation[6]{DAMTP, Centre for Mathematical Sciences, Wilberforce Road, Cambridge, CB3 0WA, UK.}

\emailAdd{laparici@ictp.it}
\emailAdd{mcicoli@ictp.it}
\emailAdd{krippendorf@th.physik.uni-bonn.de}
\emailAdd{anshumanmaharana@hri.res}
\emailAdd{muia@bo.infn.it}
\emailAdd{f.quevedo@damtp.cam.ac.uk}

\abstract{We analyse soft supersymmetry breaking in
type IIB de Sitter string vacua after moduli stabilisation, focussing on models
in which the Standard Model is sequestered from the supersymmetry breaking sources
and the spectrum of soft-terms is hierarchically smaller than the gravitino mass $m_{3/2}$.
Due to this feature, these models are compatible with gauge coupling unification and TeV scale supersymmetry with no cosmological moduli problem.
We determine the influence on soft-terms of concrete realisations of de Sitter vacua constructed from supersymmetric effective actions. One of these scenarios provides the first study of soft-terms for consistent string models embedded in a compact Calabi-Yau manifold with all moduli stabilised.
Depending on the moduli dependence of the K\"ahler metric for matter fields and on the mechanism responsible to obtain a de Sitter vacuum,
we find two scenarios for phenomenology: (i) a split-supersymmetry scenario where gaugino masses are suppressed with respect to scalar masses: $M_{1/2} \sim m_{3/2}\epsilon\ll m_0 \sim m_{3/2} \sqrt{\epsilon}\ll m_{3/2}$ for $\epsilon \sim m_{3/2}/M_P\ll 1$;
(ii) a typical MSSM scenario where all soft-terms are of the same order: $M_{1/2} \sim m_0 \sim m_{3/2}\epsilon\ll m_{3/2}$.
Background fluxes determine the numerical coefficients of the soft-terms allowing for small variations of parameters as is necessary to confront data and to interpolate between different scenarios.
We comment on different stringy origins of the $\mu$-term and potential sources of desequestering.}

\preprint{DAMTP-2014-21 \\
\phantom{a} \hfill{HRI/ST1410}}


\begin{document}

\maketitle

\bigskip

\section{Introduction}

The simplest models of low energy supersymmetry (SUSY) as a solution to the hierarchy problem
are in tension with the latest LHC results (see e.g.~\cite{Craig:2013cxa} and references therein)
which are moving the bounds for sparticle masses beyond the TeV scale. We are then either in a situation where we accept two to three orders of magnitude of tuning as still `natural', or we are at a very particular corner in the Minimal Supersymmetric Standard Model (MSSM) parameter space with less fine-tuning (e.g.~natural SUSY~\cite{Kitano:2005ew,Papucci:2011wy,Brust:2011tb}, compressed spectra~\cite{Lebedev:2005ge,LeCompte:2011cn},
RPV models~\cite{Allanach:2012vj,Evans:2012bf}), or we need alternatives to the conventional MSSM. Given this, there are various avenues to explore for addressing the electroweak hierarchy problem:
\begin{enumerate}
\item{} The simplest MSSM models (e.g.~CMSSM) need to be modified at low energies to account for particular corners in the MSSM parameter space with reduced fine-tuning. Or, one step further, extensions of the MSSM including extra matter and/or interactions at the TeV scale may relax the tuning of the MSSM (see e.g.~\cite{Ross:2012nr}).
\item The MSSM is the correct description for beyond the Standard Model (SM) physics
but the hierarchy problem is addressed by different amounts of fine-tuning through the multiverse just like the cosmological constant problem \cite{Bousso:2000xa}, where we can distinguish the following classes:
\begin{enumerate}
\item The simplest models of low energy SUSY are realised with some two to three orders magnitude of fine-tuning.
\item One just keeps the appealing features of low-energy SUSY of realising the correct dark matter density and gauge coupling unification whereas the hierarchy problem is no longer addressed. This proposal is commonly referred to as split SUSY \cite{ArkaniHamed:2004fb} where gauginos are at the TeV scale while the scalar superpartners are hierarchically heavier.
\item Dark matter and gauge coupling unification are achieved by other mechanisms and the SUSY particles are at a scale far above the electroweak scale such as an intermediate scale.
\end{enumerate}
\item One can consider alternative solutions to the hierarchy problem such as composite models or extra-dimensional models.
\end{enumerate}

Each of these scenarios has its own virtues and demerits. The first one aims at avoiding fine-tuning in the parameter space of the MSSM, but without a principle on why to favour a particular extension in a UV theory, it is in some sense a tuning in theory space which is as appealing as fine-tuning in parameter space, the others simply accept some sort of tuning.\footnote{Particular interesting corners of parameter space for soft-terms can be obtained by invoking principles such as precision gauge coupling unification~\cite{Krippendorf:2013dqa} or by identifying pattern in underlying UV theories (e.g.~realisation of natural SUSY and compressed spectra in the heterotic mini-landscape~\cite{Krippendorf:2012ir,Badziak:2012yg}).} Given this state of affairs, we are left with the unpleasant situation that at present the best argument
in favour of low-energy SUSY is that other alternatives, like large extra dimensions or composite models, are looking even worse.

This is a golden opportunity for string theoretical scenarios to play a role.
Being the only explicit scenarios that provide a UV completion of the SM,
they should be able to address the problems of the scenarios mentioned above,
provide guidance towards their explicit realisation and maybe even suggest other alternative avenues.

Consistent string theories are typically supersymmetric. Unfortunately, low-energy SUSY or the MSSM
are not a prediction of string theory and its potential discovery or lack of will not directly test string theory.
Moreover for a high string scale of order $10^{16}$~GeV (as hinted by standard MSSM unification and recent inflationary observations \cite{Ade:2014xna}),
obtaining at the same time low-energy SUSY can be a challenge for model building.
Another important feature is the string landscape which can potentially have an impact on the hierarchy problem.
These are very important issues which can impact LHC and future collider observations. They need to be addressed systematically
and within a complete string framework. This is the subject of the present article.

Fortunately progress in the understanding of SUSY breaking in string compactifications is maturing right on time to play a role.
Several scenarios in which most of the string moduli have been stabilised
with SUSY breaking and computable soft-terms have emerged~\cite{Choi:2005ge,Nilles:1997cm,Conlon:2006us,Conlon:2006wz, Lowen:2008fm, deAlwis:2009fn, Cicoli:2013rwa,Acharya:2008zi}.
Some of them are also consistent with cosmological constraints such as the cosmological moduli problem (CMP) and the realisation of de Sitter (dS) vacua.
In particular, the LARGE Volume Scenario (LVS)~\cite{Balasubramanian:2005zx}, on which we focus in this article,
allows for several of the above SUSY breaking scenarios in which soft-terms can be explicitly computed.
Moreover, LVS is an ideal framework to build globally consistent MSSM-like chiral models for explicit
Calabi-Yau (CY) compactifications with all closed string moduli stabilised~\cite{Cicoli:2011qg,Cicoli:2012vw,Cicoli:2013mpa,Cicoli:2013cha}.
It is also possible to obtain dS vacua from supersymmetric effective actions \cite{Cicoli:2012fh,Cicoli:2012vw} and the string landscape allows for a controllable fine-tuning of the cosmological constant and potentially the electroweak hierarchy problem.

\subsection{SUSY breaking in LVS}

Let us briefly summarise the main properties of LVS relevant for soft SUSY breaking:
\begin{itemize}
\item {\it Closed string moduli stabilisation:} Complex structure moduli and the dilaton
are fixed by three-form fluxes at a supersymmetric minimum. The degeneracy associated with the flux quanta leads to a landscape of vacua.
The non-vanishing value of the flux superpotential $W_0$ at the minimum breaks SUSY.
Perturbative and non-perturbative corrections to the tree-level effective action
fix the K\"ahler moduli at sizes larger than the string scale (as required to control the $\alpha'$ and $g_s$ expansions).
The Einstein-frame volume $\vo$ is exponentially large in string units: $\vo\sim e^{1/g_s}$ ($g_s$ is the string coupling).

\item{\it Hierarchy of scales:} LVS leads to a hierarchy of scales for masses and soft-terms \cite{Conlon:2005ki}.
In Planck units ($M_P$ is the reduced Planck mass), the string scale is (see Appendix~A of \cite{Conlon:2005ki} for the derivation of the exact prefactors)
\be
M_s = \frac{g_s^{1/4} M_P}{\sqrt{4\pi\vo}}\,,
\label{Ms}
\ee
the Kaluza-Klein scale is
\be
M_\KK\simeq \frac{M_P}{\sqrt{4\pi}\vo^{2/3}}\,,
\label{MKK}
\ee
and the gravitino mass is\footnote{We set the VEV of the K\"ahler potential for complex structure moduli such that $e^{{K_{\rm cs}}/2}=1$.}
\be
m_{3/2} = e^{K/2} |W|= \left(\frac{g_s^2}{2\sqrt{2\pi}}\right) \frac{W_0 M_P}{\vo} +\ldots\, ,
\label{m32}
\ee
where the dots indicate suppressed corrections in the inverse volume expansion.
Most of the moduli receive a mass of order $m_{3/2}$ except for the volume mode whose mass is
$m_\vo \simeq m_{3/2}/\sqrt{\vo}$. Hence there is a natural hierarchy of scales $M_s\gg M_\KK\gg m_{3/2}\gg m_\vo$
for the flux superpotential $W_0$ taking generic values between $1-100$.

\item {\it Bottom-up model building:} The D-brane configuration of the visible sector is localised in a particular corner
of the bulk geometry, allowing for a realisation of the bottom-up approach to string model building~\cite{Aldazabal:2000sa}.
The structure of soft-terms does not depend on the gauge theory realised in the visible sector but only on the type of D-brane configuration
(e.g. branes at singularities, D7-branes in the geometric regime) as in the modular approach to string model building.
The realisation of the visible sector on a cycle different from the one supporting non-perturbative effects allows to achieve compatibility of chirality and moduli stabilisation~\cite{Blumenhagen:2007sm,Cicoli:2011qg}.

\item {\it SUSY breaking:} Assuming a D-brane configuration that leads to the MSSM,
the effective field theory allows to analyse the structure of soft-masses.
In particular, the pattern of soft masses depends on the location and type of the MSSM D-brane construction
in the CY orientifold compactification.
If the MSSM is located at a divisor geometrically separated from the main sources of SUSY breaking in the bulk,
e.g.~on a shrinking divisor, there can be a hierarchical suppression of the soft masses below the gravitino mass and the lightest modulus~\cite{Blumenhagen:2009gk}. If the dominant source of SUSY breaking is in the proximity of the visible sector brane configuration
(as it happens if the F-term of the modulus of the cycle wrapped by the SM brane breaks SUSY),
the soft masses are of order the gravitino mass with only mild suppressions~\cite{Conlon:2005ki, Conlon:2010ji,Choi:2010gm,Shin:2011uk}.
\end{itemize}

Generically moduli masses tend to be of order the gravitino mass.
In view of the CMP which sets a lower bound on moduli masses of order $50$ TeV~\cite{Coughlan:1983ci,Banks:1993en,deCarlos:1993jw},
it is often desirable to have soft masses well below the gravitino/moduli masses
although achieving this requires a special mechanism at play. We will refer to models which have hierarchically suppressed soft masses
(not just by loop factors) as sequestered models.\footnote{A similar suppression appears also in the context of realisations of the KKLT scenario~\cite{Choi:2005ge}.}
Depending on the location of SM particles and the value of the CY volume, we distinguish three interesting LVS scenarios for SUSY breaking:
\begin{enumerate}
\item{\bf Unsequestered GUT scale string models:}
Motivated by unification, if one takes the string scale to be close to the GUT scale $10^{14}-10^{16}$ GeV, where the range in the volume captures the uncertainty about high-scale threshold corrections,
then the volume is of order $\vo\simeq 10^{3}-10^{7}$ for $g_s\simeq 0.1$.
This implies a large gravitino mass, $m_{3/2}\simeq 10^{10}-10^{14}$ GeV, i.e.~unobservable sparticles,
unless the flux superpotential is tuned to extremely small values (tuning of up to $W_0\sim 10^{-10}$)
to get TeV soft-terms.\footnote{TeV scale soft masses in this scenario would lead to light moduli
which suffer from the CMP since $m_\vo \simeq 10$ GeV.}
So the generic situation without tuning $W_0$ is that soft-terms are at an intermediate scale, roughly in the range $10^{10}-10^{14}$ GeV which is the option 2c) described earlier.
This scenario is safe from the CMP. The string landscape can in principle address the hierarchy problem.

\item{\bf Unsequestered intermediate scale strings:}
Requiring TeV scale soft-terms in an unsequestered setting leads to a volume of order $\vo\simeq 10^{14}$ for $W_0\sim 10$,
implying an intermediate string scale, $M_s\simeq 5\cdot 10^{10}$ GeV.
This scenario addresses the hierarchy problem, although unification has to work differently from the MSSM (see \cite{Aldazabal:2000sk, 1106.6039,Cicoli:2013mpa}
for concrete string examples with intermediate scale unification). Its spectrum of soft-terms at the electroweak scale has been studied in \cite{Conlon:2007xv}.
It suffers from the CMP since the volume modulus mass is slightly below 1 MeV.

\item{\bf Sequestered high scale string models:}
There can be special situations in which the soft-terms are hierarchically smaller than the gravitino mass, referred to as
sequestered scenarios~\cite{Blumenhagen:2009gk}.
In LVS this happens in configurations in which the SM degrees of freedom are localised in the extra dimensions,
such as in models where the visible sector arises from open strings on D3-branes at a singularity.
In particular, in this setup the F-term of the SM cycle vanishes and the dominant F-terms are associated with other moduli
(the volume modulus, the dilaton and other K\"ahler and complex structure moduli).
However the dominant F-terms couple very weakly to the visible sector because of their bulk separation,
and this produces a hierarchy between the soft-terms and the gravitino mass.
Typically gaugino masses are of order $M_{1/2}\simeq m_{3/2}/\vo$,
whereas scalar masses can be as suppressed as the gaugino masses or hierarchically larger by a power $\vo^{1/2}$
(leading to a split SUSY scenario in this last case).
This makes these models very attractive for phenomenology since they feature TeV scale soft-terms
and no CMP for $\vo\simeq 10^7$ and $W_0\simeq 50$ which give
$M_{1/2}\simeq 1$~TeV, $m_{3/2}\simeq 10^{10}$ GeV and $m_\vo \simeq 5\cdot 10^6$ GeV. The unification scale in these models is set by
the winding scale $M_W = 2 \pi\sqrt{\pi g_s} M_P / \vo^{1/3}$ \cite{Conlon:2009xf,Conlon:2009kt} which turns out to be of the same order of the standard GUT scale. The appearance of this hierarchical suppression of soft masses is subject to the structure of the effective supergravity. Changes to the EFT at loop or non-perturbative level (see for instance~\cite{Conlon:2010ji,Choi:2010gm, Berg:2010ha, Conlon:2011jq,Berg:2012aq}) can lead to desequestering. In Appendix~\ref{sec:deseq} we comment more explicitly on possible sources for desequestering and focus for the remainder of this paper on constructions where these desequestering effects can be absent.
\end{enumerate}

\subsection{Overview}

In this paper our focus shall be on the last of the three scenarios described above: sequestered models.
In~\cite{Blumenhagen:2009gk} it was realised that soft-terms
can potentially be sensitive to the mechanism responsible for achieving a dS minimum.
Lack of a controlled understanding of the way to get dS vacua made it difficult to present a complete analysis of the SUSY phenomenology.
Recently there has been progress in obtaining dS vacua from supersymmetric effective actions~\cite{Cicoli:2012fh,Cicoli:2012vw,Cicoli:2013mpa,Cicoli:2013cha,Krippendorf:2009zza}. In this paper we work out this dependence on the uplifting mechanism in sequestered models. As previously, we assume an MSSM spectrum from the local D-brane configuration for simplicity.
To perform the lengthy soft-term computations we have developed a code called \texttt{Large$\mathcal{V}$ol}.\footnote{\texttt{Large$\mathcal{V}$ol} is
a Mathematica Package useful to analyse the phenomenology of various type IIB supergravity theories.
It computes and minimises the scalar potential following the LVS mechanism for moduli fixing.
\texttt{Large$\vo$ol} can calculate F-terms and soft-terms generated via both supergravity and anomaly mediation.}

We explicitly compute all soft-terms for sequestered scenarios
identifying different cases depending on the mechanism to obtain dS vacua and the moduli-dependence of
the K\"ahler metric for matter fields.
Broadly, we find two classes of models: scenarios in which all soft-terms are of order $m_{3/2}/\vo$
and scenarios where gaugino masses and A-terms are of this order but scalar masses are of order $m_{3/2}/\vo^{1/2}$.
In both cases the numerical coefficients of the soft-terms are determined by background fluxes and therefore can be tuned by scanning through the landscape.
This provides an explicit mechanism for the (small) tuning that might be necessary to confront LHC data.
In the first class of models the spectrum is similar to standard MSSM spectra with soft-terms of the same order but with the potential of extra non-universal flux dependent contribution.
The second one gives a universal mini split scenario with negligible non-universalities. We leave a detailed study of the LHC phenomenology of these models to a companion article \cite{us}.

The rest of this paper is organised as follows. Sec.~\ref{slvs} contains the detailed setup that leads to sequestered LVS models and a presentation of two mechanisms to obtain dS vacua. We then compute the leading order expressions of the associated F-terms and soft-masses for these scenarios in Sec.~\ref{sec:ftermsandsoftmasses} before concluding in Sec.~\ref{sec:conclusions}. Finally in Appendix~\ref{App} we present subleading corrections to F-terms while in Appendix~\ref{sec:deseq} we comment on possible sources of desequestering.

\section{Sequestered LVS scenarios}
\label{slvs}

\subsection{General setup}
\label{generalsetup}

Let us outline a setup in type IIB CY flux compactifications with O3/O7-planes that leads to moduli stabilisation
\`a la LVS and a visible sector sequestered from SUSY breaking:

\begin{itemize}
\item The simplest LVS vacua can be obtained for a CY with negative Euler number and at least one blow-up
of a point-like singularity \cite{Cicoli:2008va}. For these manifolds the volume~$\vo$ is of Swiss-cheese type
\be
\vo = \alpha_b \tau_b^{3/2} - \sum_i \alpha_i \tau_i^{3/2}\,,
\label{eq:vol}
\ee
where $\tau_b$ denotes the overall big four-cycle volume and the $\tau_i$ denote blow-up moduli. The numerical coefficients $\alpha_{b,i}$ are determined by the CY triple intersection numbers and in what follows.\footnote{It is possible to implement LVS in CYs which have a more general volume form~\cite{Cicoli:2008va} but this does not alter the structure of soft-masses and so we do not consider these cases.}

\item The visible sector can be realised with appropriate D-brane configurations on blow-up moduli. Concrete D-brane realisations with D3/D7 branes at del Pezzo singularities can lead to interesting gauge/matter extensions of the MSSM. As we will discuss in Sec.~\ref{shrink}, the size of the associated four-cycle can shrink to zero value due to D-term stabilisation. Because of this shrinking, the F-term of the corresponding blow-up K\"ahler modulus is vanishing at leading order giving rise to a sequestered scenario.

\item In order to realise a dS vacuum one introduces further ingredients in the compactification. Here we concentrate on two options:
(i) Hidden sector matter fields on the large cycle which acquire non-zero F-terms because of D-term fixing~\cite{Cicoli:2012vw}; (ii) E(-1) instantons at a second singularity whose blow-up mode develops non-vanishing F-terms due to new dilaton-dependent non-perturbative effects \cite{Cicoli:2012fh}. These mechanisms will be discussed later in this section.
\end{itemize}
This setup has been realised in concrete CY orientifold compactifications with D3(/D7) branes at singularities~\cite{Cicoli:2012vw,Cicoli:2013mpa,Cicoli:2013cha} that satisfy all global consistency conditions (e.g.~tadpole cancellation).
The minimal setup that allows this realisation includes at least four K\"ahler moduli:
a `big' four-cycle $T_b$ controlling the size of the CY volume, a `small' blow up mode $T_s$
supporting non-perturbative effects, the visible sector cycle $T_\SM$ and its orientifold image $G$.
These last two moduli are associated to two del Pezzo divisors which collapse to zero size due to D-term fixing\footnote{The positivity of soft scalar masses for visible sector fields fixes all remaining flat-directions after D-term stabilisation \cite{Cicoli:2012vw}.}
and are exchanged by the orientifold involution.
This setup leads to $h^{1,1}_+ = 3$ and $h^{1,1}_- = 1$ with the following K\"ahler moduli:
\be
T_b = \tau_b + i \psi_b\,, \quad T_s = \tau_s + i \psi_s\,, \quad T_\SM = \tau_\SM + i \psi_\SM\,, \quad G = b + i c\,,
\label{eq:cycles}
\ee
where $\tau_b$, $\tau_s$ and $\tau_\SM\to 0$ are divisor volumes, the $\psi$'s are axions given by the reduction of $C_4$ on each of the relevant four-cycles,
whereas $b$ and $c$ are respectively the reduction of $B_2$ and $C_2$ on the two-cycle dual to the shrinking one.
The CY volume $\vo$ is a function of the K\"ahler moduli which takes the same form as in (\ref{eq:vol}).

\subsubsection{${\cal N} = 1$ supergravity effective field theory}

In this section we review the low energy effective action relevant for our construction
in the language of 4D ${\cal N}=1$ supergravity.
We take the superpotential of the following form
\be
W= W_{\rm flux}(U,S) + A_s(U, S)\, e^{- a_s T_s} +  W_\dS+W_{\rm matter}\,.
\label{W}
\ee
$W_{\rm flux}$ is the standard flux-generated superpotential~\cite{Gukov:1999ya}.
The second term incorporates non-perturbative effects on the `small' blow-up cycle which can arise from gaugino condensation or ED3-instantons.
The type of D-brane configuration determines the coefficient $a_s$ and the prefactor $A_s(U,S)$ depends on both complex structure moduli $U$ and the dilaton $S$ whose real part $s$ sets the string coupling: $\langle s\rangle =g_s^{-1}$.\footnote{The dependence on $S$ and $U$-moduli is structurally different, i.e.~the dependence on the dilaton is generated when including the backreaction of sources and warping on the geometry \cite{Baumann:2006th}.}
The term $W_\dS$ involves the contribution from the mechanism used to obtain a dS vacuum (see Sec.~\ref{dsv})
while $W_{\rm matter}$ is the visible sector superpotential
\be
W_{\rm matter} = \mu(M) H_u H_d + \frac 16 Y_{\alpha\beta\gamma}(M) C^\alpha C^\beta C^\gamma + \cdots\,,
\label{Wmatter}
\ee
where we denoted the moduli as $M$ and the MSSM superfields as $C^\alpha$. Moreover, the dots refer to higher dimensional operators.
We also separated the two Higgs doublets $H_u$ and $H_d$ from the rest of matter fields in the moduli-dependent $\mu$-term.
Because of the holomorphicity of $W$ and the perturbative shift symmetry of the axionic components of the K\"ahler moduli,
the Yukawa couplings and the $\mu$-term can depend only on $S$ and $U$ at the perturbative level with the $T$-moduli appearing only non-perturbatively. We discuss this dependence in more detail in Sec.~\ref{sec:ftermsandsoftmasses} and Appendix~\ref{sec:deseq}.

As motivated in~\cite{0810.5660,Blumenhagen:2009gk}, we assume the following form of the K\"ahler potential which describes the regime for the visible sector near the singularity
\be
\label{generalk}
K = - 2 \ln\left(\vo + \frac{\hat\xi}{2}\right) - \ln(2s) + \lambda_\SM \frac{\tau_\SM^2}{\vo} + \lambda_b \frac{b^2}{\vo}
+ K_\dS + K_{\rm cs}(U) + K_{\rm matter}\, ,
\ee
where $\hat\xi\equiv \xi s^{3/2}$, the $\lambda$'s are $\mc{O}(1)$ coefficients, $K_{\rm cs}(U)$ is the tree-level K\"ahler potential for complex structure moduli and $K_\dS$ encodes the dependence on the sector responsible
for obtaining a dS vacuum (see Sec.~\ref{dsv}).
The matter K\"ahler potential $K_{\rm matter}$ is taken to be
\be
K_{\rm matter} = \tilde{K}_{\alpha}(M, \overline{M}) \overline{C}^{\overline{\alpha}} C^{\alpha} + [Z(M, \overline{M}) H_u H_d + \text{h.c.}]\,.
\label{mm}
\ee
We assume at this stage that the matter metric is flavour diagonal beyond the leading order structure which was highlighted in~\cite{Conlon:2007dw}.\footnote{Subleading flavour off-diagonal entries which can in principle appear~\cite{Camara:2013fta} are taken to be absent. This is motivated by the appearance of additional anomalous $U(1)$ symmetries in D-brane models, in particular also in the context of del Pezzo singularities~\cite{1106.6039}.} The only exception is that we allow for the Higgs bilinear to appear in $K_{\rm matter}$
which we parameterise with the function $Z$.
Note that $\tilde{K}_\alpha$ is the matter metric for the visible sector which we will parameterise as~\cite{Blumenhagen:2009gk}
\be
\tilde{K}_{\alpha} = \frac{f_\alpha(U,S)}{\cv^{2/3}} \left(1 - c_s \frac{\hat\xi}{\cv}
+ \tilde{K}_\dS + c_\SM \tau_\SM^p + c_b b^p\right), \qquad p > 0\,,
\label{mattermetric}
\ee
where we have used $\tilde{K}_\dS$ to parameterise the dependence on the
dS mechanism (details will be given in Sec.~\ref{soft}). The $c$'s are taken as constants for simplicity while $p$ is taken to be positive
in order to have a well-behaved metric in the singular limit $b,\tau_\SM \to 0$.
As they can in principle depend on $U$ and $S$, we comment in due course on the influence on the soft-terms of such a dependence.
The appearance of the Higgs bilinear and its potential parametrisation are discussed in Sec.~\ref{bmusection} when we analyse the $\mu$-term in this scenario. In general the functions $f_\alpha(U,S)$ could be non-universal. Such non-universality can have interesting phenomenological implications (e.g.~mass hierarchies among families of sfermion masses needed for a realisation of natural SUSY).
As we are interested in soft-terms arising for D-branes at singularities, we take the gauge kinetic function to be
\be
f_a=\delta_a S+ \kappa_a \,T_\SM\, ,
\label{eq:gaugekineticf}
\ee
where $\delta_a$ are universal constants for $\mathbb{Z}_n$ singularities but can be non-universal for more general singularities.

\subsection{Moduli stabilisation}

As outlined earlier in this section, we stabilise the moduli following the LVS procedure.
The complex structure moduli and the dilaton are fixed at tree-level by background fluxes while the K\"ahler moduli are fixed using higher order corrections to the effective action \cite{Cicoli:2013cha}.

\subsubsection{D-term stabilisation}
\label{shrink}

The K\"ahler moduli where the visible sector D-brane configuration is located are stabilised using D-terms which are the leading order contribution to the potential. Remaining flat directions are stabilised using subleading F-term contributions. To set the notation, let us review D-term stabilisation \cite{Cicoli:2012vw}. The moduli $T_\SM$ and $G$ are charged under two anomalous $U(1)$ symmetries with charges $q_1$ and $q_2$. The corresponding D-term potential reads
\be
V_D = \frac{1}{2{\rm Re}(f_1)} \left(\sum_\alpha q_{1 \alpha} \frac{\partial K}{\partial C^\alpha} C^\alpha - \xi_1\right)^2
+ \frac{1}{2{\rm Re}(f_2)} \left(\sum_\alpha q_{2 \alpha} \frac{\partial K}{\partial C^\alpha} C^\alpha - \xi_2\right)^2\,,
\label{dtermpot}
\ee
where $f_1$ and $f_2$ are the gauge kinetic functions of the two $U(1)$s.
The Fayet-Iliopoulos~(FI) terms are given by (see appendix of \cite{Cicoli:2011yh} for the exact numerical factors)
\begin{eqnarray}
&& \xi_1 = - \frac{q_1}{4\pi} \frac{\partial K}{\partial T_\SM} = - \frac{q_1  \lambda_\SM}{4\pi} \frac{\tau_\SM}{\vo}\,, \label{xiSM} \\
&& \xi_2 = - \frac{q_2}{4\pi} \frac{\partial K}{\partial G} = - \frac{q_2 \lambda_b}{4\pi} \frac{b}{\vo}\, .
\end{eqnarray}
The vanishing D-term condition fixes therefore $\tau_\SM$ and $b$ in terms of visible sector matter fields.
The remaining flat directions are fixed by subleading F-term contributions which give vanishing VEVs to the $C^\alpha$ if they develop non-tachyonic soft masses from SUSY breaking \cite{Cicoli:2012vw}.\footnote{If the soft scalar masses of some $C^\alpha$ are tachyonic,
they develop non-zero VEVs (which could be phenomenologically allowed for some SM singlets) that, in turn, induce non-zero FI-terms \cite{Cicoli:2013cha}.
However $\tau_\SM$ and $b$ would still be fixed in the singular regime since their VEVs would be volume-suppressed \cite{Cicoli:2013cha}.}
Hence the D-term potential (\ref{dtermpot}) vanishes in the vacuum since it is fixed to a supersymmetric minimum at $\xi_1 = \xi_2 = 0$. This corresponds to the singular limit $\tau_\SM = b = 0$. In turn, the axions $\psi_\SM$ and $c$ are eaten up by the two $U(1)$ gauge bosons in the process of anomaly cancellation.

\subsubsection{F-term stabilisation}

Analysing the F-term scalar potential in an inverse volume expansion, one finds that the leading contribution scales as $\vo^{-2}$.
This is generated by the flux superpotential $W_{\rm flux}$ and is positive semi-definite
\be
\label{vddilaton}
V_{\mc{O}({\vo}^{-2})} = \frac{1}{2 s\vo^2} \left.\left[4 s^2 |D_S W_{\rm flux}|^2
+ K^{U \overline{U}} D_U W_{\rm flux} D_{\overline{U}} \overline{W}_{\rm flux}\right]\right|_{\xi=0},
\ee
where the subscript $\xi = 0$ denotes that $\alpha'$ corrections can be neglected at this level of approximation.
This potential fixes the dilaton and the complex structure moduli at a supersymmetric minimum located at
\be
\label{susyminim}
\left.D_S W_{\rm flux}\right|_{\xi  = 0} = 0\ , \qquad D_U W_{\rm flux} \big{|}_{\xi = 0} = 0\ , \qquad \langle W_{\rm flux}\rangle\equiv W_0\ .
\ee
The K\"ahler moduli are stabilised using $\alpha'$ corrections to $K$ (\ref{generalk})
and non-perturbative corrections to $W$ (\ref{W}) which give rise to $\mc{O}(\vo^{-3})$ contributions
to the scalar potential\footnote{We have already fixed the axion $\psi_s$ at $a_s\langle\psi_s\rangle=\pi$.
The axion $\psi_b$ associated to the large cycle can develop a potential only via $T_b$-dependent non-perturbative effects.}
\be
\label{v0}
V_{\mc{O}({\vo}^{-3})} = \frac{1}{2 s} \left[\frac{8}{3}(a_s A_s)^2 \sqrt{\tau_s} \frac{e^{- 2 a_s \tau_s}}{\vo}
- 4 a_s A_s W_0 \tau_s \frac{e^{-a_s \tau_s}}{\vo^2 } + \frac{3\hat\xi W_0^2}{4\vo^3}\right]\, .
\ee
This potential admits an AdS global minimum which breaks SUSY. Minimisation with respect to $\tau_s$ yields
\be
e^{- a_s \tau_s} = \frac{3 \sqrt{\tau_s} W_0}{4 a_s A_s \vo} \frac{\left(1-4\epsilon_s\right)}{\left(1-\epsilon_s\right)}\qquad\text{with}\qquad \epsilon_s \equiv\frac{1}{4 a_s \tau_s}\sim\mc{O}\left(\frac{1}{\ln\vo}\right)\ll 1\,.
\label{eataus}
\ee
On the other hand, minimisation with respect to $\tau_b$ gives
\be
\tau_s^{3/2} = \frac{\hat\xi}{2} \left[1 + f_\dS(\epsilon_s) \right]\,,
\label{tausms}
\ee
where $f_\dS$ is a subdominant function of $\epsilon_s$ which depends on the particular mechanism used to obtain a dS vacuum (see Appendix~\ref{App}).
The relation (\ref{tausms}) implies that at the minimum (neglecting $f_\dS$)
\be
\hat\xi\simeq \frac{1}{4\left(a_s \epsilon_s\right)^{3/2}}\sim\mc{O}\left[\left(\ln\vo\right)^{3/2}\right] \gg 1\,.
\label{tausms2}
\ee
Given that the potential (\ref{v0}) depends on $S$ and $U$ (via $A_s(U,S)$ and $s$-dependent $\alpha'$ effects),
the minimum (\ref{susyminim}) is slightly shifted from its supersymmetric locus. This shift is fundamental for the soft-term computation
in sequestered scenarios since non-vanishing F-terms of $U$ and $S$ at subleading order
can actually provide the main contribution to soft-terms~\cite{Blumenhagen:2009gk}.

\subsubsection{Shift in the dilaton and complex structure minimum}
\label{shiftsection}

Let us try to estimate the shift of $S$ and $U$ from their supersymmetric minimum (\ref{susyminim}) because of $\alpha'$
and non-perturbative effects. The K\"ahler covariant derivative of the total superpotential evaluated at the minimum (\ref{eataus}) and (\ref{tausms})
reads (we neglect $\mc{O}(\epsilon_s)$ effects)
\be
D_S W \simeq \left.D_S W_{\rm flux}\right|_{\xi = 0} - \frac{3\hat\xi W_0}{4s\vo}
\left[1 + \epsilon_s s\partial_s \ln A_s(U,S)\right]\,.
\ee
Since we do not know the functional dependence of $A_s(U,S)$ and since the $W_\dS$ term in (\ref{W})
can also potentially shift the dilaton minimum, it is not
possible for us to compute this shift explicitly. We will parameterise it by using the parameter $\omega_S(U,S)$ defined as
\be
\label{dilmin}
D_S W = - \frac{3 \omega_S(U,S)}{4} \frac{\hat\xi W_0}{s\vo}\,.
\ee
The dependence of $A_s(U, S)$ on the complex structure moduli is also responsible for shifting the $U$-moduli
from their supersymmetric minimum. After imposing the minimisation conditions, the total $D_U W$ looks like (denoting $u\equiv {\rm Re}(U)$)
\be
D_U W \simeq \left.D_U W_{\rm flux}\right|_{\xi = 0} - \frac{3\hat\xi W_0}{4\vo}
\epsilon_s \partial_u\left[K_{\rm cs}(U)+  \ln A_s(U,S)\right]\,,
\ee
and so we can parameterise this shift by $\omega_{U_i}(U,S)$ as
\be
\label{csmin}
D_{U_i} W = - \frac{3 \omega_{U_i} (U,S)}{4} \frac{\hat\xi W_0}{s\vo}
\qquad\Rightarrow\qquad D_{U_i} W =\frac{\omega_{U_i} (U,S)}{\omega_S(U,S)} D_S W\sim \mc{O}(\vo^{-1})\,,
\ee
where both $\omega_S$ and $\omega_{U_i}$ are expected to be $\mc{O}(1)$ functions of $S$ and $U$. Note that both functions $\omega_{S,U_i}$ depend also on the dS mechanism.

\subsection{Scenarios for de Sitter vacua}
\label{dsv}

In this section we review two mechanisms which can lead to dS vacua in LVS.

\subsubsection{Case 1: dS vacua from hidden matter fields}
\label{matteruplift}

\begin{figure}[t]
\begin{center}
\includegraphics[width=0.4\textwidth, angle=270]{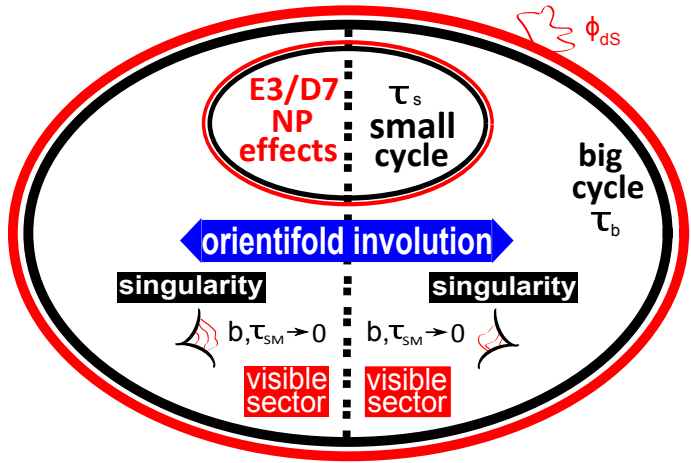}
\caption{Pictorial sketch of our CY setup for dS vacua from hidden matter fields.} \label{Fig1}
\end{center}
\end{figure}

In the LVS setting, dS vacua can arise if some hidden matter fields
acquire non-vanishing F-terms which provide a positive definite contribution to the scalar potential \cite{Cicoli:2012vw}.
The models constructed in \cite{Cicoli:2012vw} provide globally consistent explicit examples
of string models with a semi-realistic visible sector, moduli stabilisation and a positive cosmological constant
(see Fig.~\ref{Fig1} for a pictorial sketch of this setup).

Generically, the choice of $B_2$ which cancels the Freed-Witten anomaly on the small cycle $T_s$,
leads to non-vanishing gauge fluxes on the big cycle $T_b$. As a consequence, $T_b$ acquires a non-zero $U(1)$-charge $q_b$
generating a moduli-dependent FI-term. The D-term potential becomes
(focusing for simplicity on a single matter field $\phi_\dS$ with K\"ahler metric $K_\dS= s^{-1}|\phi_\dS|^2$ \cite{Conlon:2006tj, Aparicio:2008wh} and $U(1)$-charge $q_\phi$)
\be
\label{dmatter}
V_D = \frac{1}{2{\rm Re}(f_b)} \left(\frac{q_\phi}{s} |\phi_\dS|^2 - \xi_b\right)^2\,,
\ee
where $f_b = T_b$ (neglecting $S$-dependent flux corrections) and the FI-term is given by
\be
\xi_b = - \frac{q_b}{4\pi} \frac{\partial K}{\partial T_b} = \frac{3 q_b}{8\pi}\frac{1}{\vo^{2/3}}\,,
\ee
Therefore the total scalar potential takes the form
\be
V_{\rm tot} = V_D + V_F =  \frac{1}{2\vo^{2/3}} \left(\frac{q_\phi}{s}  |\phi_\dS|^2 - \frac{3 q_b}{8\pi\vo^{2/3}}\right)^2
+ \frac{1}{s} m_{3/2}^2 |\phi_\dS|^2 + V_{\mc{O}(\vo^{-3})}\,,
\label{Vtot}
\ee
where $m_{3/2}$ is the gravitino mass as in eq.~(\ref{m32}) and $V_{\mc{O}(\vo^{-3})}$ is given in (\ref{v0}).
If the two $U(1)$-charges $q_\phi$ and $q_b$ have the same sign,
$\phi_\dS$ develops a non-vanishing VEV
\be
\label{minphi}
\frac{q_\phi}{s} |\phi_\dS|^2 = \xi_b - \frac{m_{3/2}^2 \vo^{2/3}}{q_\phi}\,.
\ee
Substituting this VEV in (\ref{Vtot}) we obtain
\be
V_{\rm tot} = V_{D,0}+\frac{3 q_b}{16\pi q_\phi} \frac{W_0^2}{s  \vo^{8/3}}+V_{\mc{O}(\vo^{-3})}\,,
\label{VTOT}
\ee
where the new positive contribution can lead to an LVS dS vacuum
while the D-term potential gives rise only to a subleading effect of order
\be
V_{D,0} = \frac{m_{3/2}^4 \vo^{2/3}}{2 q_\phi^2}\sim \mc{O}\left(\vo^{-10/3}\right)\,.
\label{VD0ds1}
\ee
Following \cite{Cicoli:2012vw}, we can minimise the total scalar potential (\ref{VTOT}) with respect to $\tau_s$ and $\vo$,
finding the following value of the vacuum energy (neglecting the subleading effect of $V_{D,0}$)
\be
\langle V_{\rm tot} \rangle \simeq \frac{3 W_0^2}{8s a_s^{3/2}\langle\vo\rangle^3}\left[\delta\, \vo^{1/3}
-\sqrt{\ln\left(\frac{\langle\vo\rangle}{W_0}\right)}\right]\,,
\label{VTOTfin}
\ee
where
\be
\delta = \frac{1}{18\pi}\frac{q_b\,a_s^{3/2}}{q_\phi} \simeq 0.02 \left(\frac{q_b\,a_s^{3/2}}{q_\phi}\right)\,.
\ee
A cancellation of the vacuum energy at $\mc{O}(\vo^{-3})$ requires therefore to tune $W_0$ so that
(a subleading tuning is needed to cancel also $V_{D,0}$)
\be
\left[\ln\left(\frac{\langle\vo\rangle}{W_0}\right)\right]^{3/2}= \delta^3\, \langle\vo\rangle \sim 5\cdot 10^{-6}\, \langle\vo\rangle\qquad\Leftrightarrow\qquad
|\phi_\dS|^2 = \frac{27 s}{4 a_s^{3/2}\vo}\sqrt{\ln\left(\frac{\langle\vo\rangle}{W_0}\right)}\sim \frac{1}{\vo\sqrt{\epsilon_s}}\,.
\label{CCconddS1}
\ee
For natural $\mc{O}(1)$ values of all underlying parameters, this relation gives a minimum for $\vo$ at order $10^6 - 10^7$ (see~\cite{Cicoli:2012vw}).

\subsubsection{Case 2: dS vacua from non-perturbative effects at singularities}
\label{dilatonuplift}

\begin{figure}[t]
\begin{center}
\includegraphics[width=0.45\textwidth]{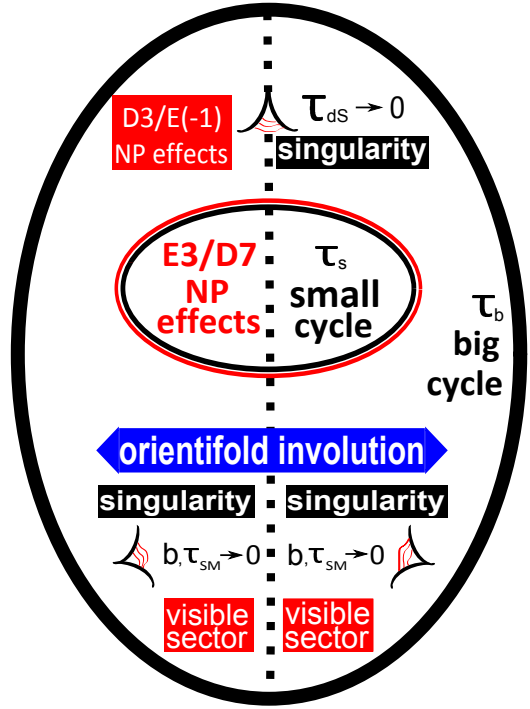}
\caption{Pictorial view of our CY setup for dS vacua from non-perturbative effects at singularities.} \label{Fig2}
\end{center}
\end{figure}

Reference \cite{Cicoli:2012fh} provided a novel method for obtaining LVS dS vacua
(see Fig.~\ref{Fig2} for a pictorial sketch of this setup).
The additional contribution to the scalar potential needed to achieve a positive cosmological constant arises
from non-perturbative effects at singularities (like gaugino condensation on spacetime filling D3-branes or E(-1) instantons).
These effects generate a new contribution to the superpotential (\ref{W}) of the form
\be
W_\dS = A_\dS(U, S) \,e^{-a_\dS (S + \kappa_\dS T_\dS)}\,.
\label{Wds2}
\ee
Because of the presence of an additional K\"ahler modulus, the K\"ahler potential (\ref{generalk}) has to be supplemented with
\be
\label{dk2}
K_\dS =  \lambda_\dS\frac{\tau_\dS^2}{\vo}\,,
\ee
with $\tau_\dS={\rm Re}(T_\dS)$. This blow-up mode can be fixed in the singular regime by minimising the hidden sector D-term potential
(focusing for simplicity on canonically normalised hidden fields $\phi_{{\rm h},i}$ with charges $q_{{\rm h},i}$ under an anomalous $U(1)$)
\be
\label{DtermPot1}
V_D = \frac{1}{2{\rm Re}(f_{\rm h})} \left(\sum_i q_{{\rm h},i} |\phi_{{\rm h},i}|^2 - \xi_{\rm h}\right)^2\,,
\ee
where $f_{\rm h} = S$ (neglecting $T_\dS$-dependent corrections) and the FI-term is given by
($q_\dS$ is the $U(1)$-charge of $T_\dS$ and from now on we set $\lambda_\dS=1$ for simplicity)
\be
\xi_{\rm h} = - \frac{q_\dS}{4\pi} \frac{\partial K}{\partial T_\dS} = - \frac{q_\dS}{4\pi} \frac{\tau_\dS}{\vo}\,.
\ee
In fact, the total scalar potential takes the leading order form (after fixing the axionic phase of $T_\dS$) \cite{Cicoli:2012fh}
\be
V_{\rm tot} = \frac{1}{2s} \left(\sum_i q_{{\rm h},i} |\phi_{{\rm h},i}|^2 + \frac{q_\dS}{4\pi} \frac{\tau_\dS}{\vo}\right)^2
+ \frac{\left(\kappa_\dS a_\dS A_\dS\right)^2}{s}\frac{e^{-2 a_\dS \left(s+\kappa_\dS\tau_\dS\right)}}{\vo} + V_{\mc{O}(\vo^{-3})}\,,
\label{Vtot1}
\ee
where the second term comes from the new superpotential (\ref{Wds2}) and $V_{\mc{O}(\vo^{-3})}$ is given in~(\ref{v0}).
Minimisation with respect to $\tau_\dS$ gives
\be
\frac{q_\dS}{4\pi} \frac{\tau_\dS}{\vo}=- \sum_i q_{{\rm h},i} |\phi_{{\rm h},i}|^2
+ \frac{a_\dS \kappa_\dS}{q_\dS}\left(\kappa_\dS a_\dS A_\dS\right)^2 e^{-2 a_\dS s}\,.
\label{tdsVEV}
\ee
Assuming that model-dependent contributions from F-terms of hidden matter fields fix some $\phi_{{\rm h}, i}$ at non-zero VEVs
such that $\langle\sum_i q_{{\rm hid},i} |\phi_{{\rm hid},i}|^2\rangle = 0$ but $A_\dS\neq 0$,\footnote{In order to make $W_\dS$ gauge invariant,
$A_\dS$ has to depend on the $\phi_{{\rm h},i}$ which can develop non-zero VEVs for appropriate hidden field F-term contributions, giving $A_\dS\neq 0$ with $\tau_\dS$ in the singular regime \cite{Cicoli:2012fh}.} and substituting the VEV (\ref{tdsVEV}) in (\ref{Vtot1}) we obtain at leading order
\be
V_{\rm tot} = V_{D,0}+ \frac{(\kappa_\dS a_\dS A_\dS)^2}{s} \frac{e^{-2 a_\dS s}}{\vo} + V_{\mc{O}(\vo^{-3})}\,.
\ee
Given that the dilaton is fixed by a ratio of flux quanta,
the extra positive-definite contribution can easily be tuned to obtain a dS minimum. Following \cite{Cicoli:2012fh}, a cancellation of the vacuum energy
at $\mc{O}(\vo^{-3})$ requires to tune 3-form fluxes such that
\be
\left(\frac{\kappa_\dS a_\dS A_\dS}{W_0}\right)^2 \,e^{-2 a_\dS s}
= \frac{9}{32} \frac{\epsilon_s \hat\xi}{\vo^2}\,.
\label{CCconddS2}
\ee
On the other hand, the D-term potential gives rise only to a subleading effect of order
\be
V_{D,0} =\frac{1}{2s} \left(\frac{a_\dS \kappa_\dS}{q_\dS}\right)^2\left(\kappa_\dS a_\dS A_\dS\right)^4 e^{-4 a_\dS s}\sim \mc{O}\left(\vo^{-4}\right)\,.
\label{VD0ds2}
\ee

\section{F-terms and soft-terms}
\label{sec:ftermsandsoftmasses}

In this section we list the leading order contributions to the F-terms relevant for the computation of all soft-terms (see Appendix~\ref{App} for the structure of subleading corrections to the F-terms). After defining our parametrisation for the K\"ahler matter metric in local and ultra-local scenarios, we then calculate the soft-terms.

\subsection{Summary of F-terms}
\label{fterms}

The general supergravity expression for an F-term is \cite{Kaplunovsky:1993rd, Brignole:1993dj}
\be
F^I = e^{K/2} K^{I \overline{J}} D_{\overline{J}} W\,.
\ee
The exact expressions for the F-terms are rather complicated. Considerable simplifications occur
if we perform an expansion in $\vo^{-1}$ and $\epsilon_s$. We also factor out the gravitino mass which is given by the following expression
\be
m_{3/2}=e^{K/2} |W|= \left(\frac{g_s^2 M_P}{2\sqrt{2\pi}}\right)\frac{W_0}{\vo} \left[ 1-\frac{\hat{\xi}}{2\vo}
\left(1+3 y_\dS \epsilon_s+\mc{O}(\epsilon_s^2)\right)+\mc{O}\left(\frac{1}{\vo^2}\right)\right]\,,
\label{expandm32}
\ee
where $y_\dS =1$ for the dS case 1 of Sec.~\ref{matteruplift} while $y_\dS = 1+\frac{\sqrt{2} a_s^{3/4}}{\kappa_\dS a_\dS}$ for the dS case 2 of Sec.~\ref{dilatonuplift}. The leading order F-terms for $T_b$ and $T_s$ turn out to be
(we show the first subleading correction only for $F^{T_b}$ since its dominant term
does not contribute to the soft-term because of the no-scale structure)
\be
\frac{F^{T_b}}{\tau_b} \simeq - 2 m_{3/2} \left(1+\frac{x_\dS}{a_s^{3/2} \vo\sqrt{\epsilon_s}}\right)\,,
\qquad
\frac{F^{T_s}}{\tau_s} \simeq - 6   m_{3/2} \epsilon_s\,,
\label{FTb}
\ee
where $x_\dS = - 45/16$ for the dS case 1 of Sec.~\ref{matteruplift} while $x_\dS \sim \mc{O}(1/\vo)$ for the dS case 2 of Sec.~\ref{dilatonuplift}
(see Appendix \ref{App}).
Because of the shift from their supersymmetric minimum, also $S$ and $U$ develop non-vanishing F-terms whose leading order
expressions are
\be
\label{ds}
\frac{F^S}{s} \simeq \frac{3 \omega'_S(U,S) }{8 a_s^{3/2}}  \frac{m_{3/2}}{\vo\epsilon_s^{3/2}}\,, \qquad
F^{U_i} \simeq - \frac{K^{U_i \overline{U}_{\overline{j}}}}{2 s^2}\frac{\omega_{\overline{U}_{\overline{j}}}(U,S)}{\omega'_S(U,S)}F^S\equiv \beta^{U_i}(U,S)F^S\,,
\ee
where $\omega'_S(U,S) \equiv 3 - 2 \omega_S(U,S)$ with $\omega_S$ as defined in (\ref{dilmin}) and $\beta^{U_i}$
are unknown $\mc{O}(1)$ functions of $U$ and $S$.
Additional non-zero F-terms are associated to fields responsible to achieve a dS solution.
For the dS case 1 of Sec.~\ref{matteruplift} there is an F-term associated to $\phi_\dS$
\be
\frac{F^{\phi_\dS}}{\phi_\dS} \simeq m_{3/2}\,,
\ee
with $\phi_\dS$ given in (\ref{CCconddS1}) (up to an irrelevant phase). On the other hand,
in the dS case 2 of Sec.~\ref{dilatonuplift} the blow-up mode $T_\dS$ has a non-vanishing F-term
(using the condition (\ref{CCconddS2}))
\be
F^{T_\dS} \simeq \frac{3}{4\sqrt{2}a_s^{3/4}}\frac{m_{3/2}}{\epsilon_s^{1/4}} \,.
\ee
Finally, the F-terms associated to the MSSM cycles $T_\SM$ and $G$ vanish:
\be
F^G = F^{T_\SM} = 0\, .
\ee
This result is crucial for sequestering since the dominant F-terms are then associated with moduli which couple weakly to the visible sector via Planck-suppressed interactions.\footnote{$T_\SM$ and $G$ can develop non-zero F-terms only in the presence of tachyonic scalar masses \cite{Cicoli:2013cha}. However, also in this case, their contribution to soft-terms turns out to be negligible.}

\subsection{Local and ultra-local scenarios}
\label{soft}

Our analysis of soft-terms will distinguish between two classes of models: local and ultra-local.
To motivate this classification we discuss the constraints (along the lines of~\cite{Conlon:2006tj}) that locality imposes on the moduli dependence
of the physical Yukawa couplings $\hat{Y}_{\alpha \beta \gamma}$
\be
\label{yuk}
\hat{Y}_{\alpha \beta \gamma} =e^{K/2}\,\frac{Y_{\alpha \beta \gamma}(U,S)}{\sqrt {\tilde{K}_\alpha \tilde{K}_\beta \tilde{K}_\gamma} }\, ,
\ee
where $Y_{\alpha \beta \gamma}(U,S)$ are the holomorphic Yukawas which do not depend on the K\"ahler moduli
due to the holomorphicity of $W$ and the perturbative shift symmetry of the $T$-axions. The K\"ahler potential $K$ is given in (\ref{generalk})
whereas the matter metrics $\tilde{K}_\alpha$, $\tilde{K}_\beta$ and $\tilde{K}_\gamma$ are given in (\ref{mattermetric}).
Since the physical Yukawas are determined by local interactions of open string
degrees of freedom, one expects at leading order that their strength is insensitive
to the overall volume of the compactification. Thus for (\ref{yuk}) to yield a result independent of $\vo$
one needs at leading order in an inverse volume expansion (for $\tau_\SM = b= C^\alpha= 0$)
\be
\label{ul}
\tilde{K}_\alpha = h_\alpha(U,S)\,e^{K/3} \simeq  \frac{h_\alpha(U,S)\,e^{K_{\rm cs}/3}}{(2s)^{1/3}\vo^{2/3}}
\left(1- \frac{\hat\xi}{3\vo}  + \frac 13 K_\dS\right)\,,
\ee
where $h_\alpha(U,S)$ is an unknown function of $U$ and $S$ and in the approximation we focus on the first subleading order corrections, e.g.~neglecting higher order corrections of $\mc{O}\left(1/\vo^{8/3}\right)$. Note
that this result has the same volume scaling of our formula for the matter metric (\ref{mattermetric}) which for $\tau_\SM = b= 0$ reduces to
\be
\tilde{K}_{\alpha} = \frac{f_\alpha(U,S)}{\vo^{2/3}} \left(1 - c_s \frac{\hat\xi}{\vo}
+ \tilde{K}_\dS\right)\equiv f_\alpha(U,S)\tilde{K}\,.
\label{newKtilde}
\ee
As found in \cite{Blumenhagen:2009gk}, our soft-term computation is sensitive to the form of $\tilde{K}_\alpha-$
beyond leading order in a $\vo^{-1}$ expansion. There is no reason
to expect that (\ref{ul}) still holds beyond leading order
since we cannot use locality to fix the form of $\tilde{K}_\alpha$
(although there is some evidence from perturbative string computations \cite{Conlon:2011jq}).
It was noted in~\cite{Blumenhagen:2009gk} that the relation (\ref{ul}) has
interesting implications for the soft-terms. Guided by this, we organise our analysis of models into two classes of phenomenological models:
\begin{itemize}
\item {\it Local}: We call a scenario `local' if (\ref{ul}) holds only to leading order in $\vo^{-1}$;
\item {\it Ultra-local}: We call a scenario `ultra-local' if (\ref{ul}) holds exactly.
\end{itemize}
If we parameterise $\tilde{K}_\dS$ as $\tilde{K}_\dS  =  c_\dS K_\dS$,
comparing (\ref{ul}) with (\ref{newKtilde}), we find that in the ultra-local case
\be
f_{\alpha}(U,S) = \frac{h_\alpha(U,S)\,e^{K_{\rm cs}/3}}{(2s)^{1/3}}\qquad \text{and}\qquad c_s=c_\dS=\frac 13\,.
\label{fdef}
\ee
Subleading deviations from the approximation in~(\ref{ul}) can be accounted for by small changes in $c_s$ and $c_\dS$ at the appropriate subleading order.

\subsection{Soft-terms}

We now proceed to compute all soft-terms distinguishing between
ultra-local and local scenarios. Throughout this section we work to
leading order in $\vo^{-1}$ and $\epsilon_s$.

\subsubsection{Gaugino masses}

The general expression for gaugino masses in gravity mediation is
\be
\label{gauginomass}
M_a = \frac{1}{2 {\rm Re}\left(f_a\right)} F^I \partial_I f_a\, ,
\ee
where $f_a=\delta_a S+ \kappa_a \,T_\SM$ is the gauge kinetic function as in~\eqref{eq:gaugekineticf}. As $F^{T_\SM}=0$, we obtain universal gaugino masses,
$M_1=M_2=M_3=M_{1/2}$, which are generated by the dilaton F-term. Potential non-universalities can arise through anomaly mediated contributions which turn out to be subleading (see Appendix~\ref{anosec} for more details).
The leading order expression for the gaugino masses is
\be
M_{1/2} = \frac{F^S}{2 s} \simeq \frac{3 \omega'_S(U,S)}{16 a_s^{3/2}}  \frac{m_{3/2}}{\vo\epsilon_s^{3/2}}
\sim \mc{O}\left(m_{3/2}\frac{\left(\ln\vo\right)^{3/2}}{\vo}\right)\ll m_{3/2}\,.
\label{ggmm}
\ee
Note that this leading order result depends on the shift of the dilaton minimum
induced by $\alpha'$ and non-perturbative effects (see Sec.~\ref{shiftsection}).
In this paper we neglect possible phases of gaugino masses. We will return to this question in the context of the low-energy analysis of soft-terms~\cite{us}.

\subsubsection{Scalar masses}
\label{scalarsection}

Scalar masses in gravity mediation receive both F- and D-term contributions. Let us study them separately, presenting their leading order expressions.\\

\noindent \textbf{F-term contributions}
\medskip\\
Assuming a diagonal K\"ahler matter metric as in~(\ref{mm}), the general expression for the F-term contributions to scalar masses in gravity mediation is \cite{Brignole:1993dj}
\be
\label{genscalF}
\left.m_\alpha^2\right|_F = m_{3/2}^2 - F^I \overline{F}^{\overline{J}} \partial_I \partial_{\overline{J}} \ln \tilde{K}_\alpha\,.
\ee

\emph{Local limit}: In the local limit we obtain universal scalar masses, $m_\alpha^2=m_0^2$ $\forall \alpha$, where
\be
\left.m_0^2\right|_F \simeq m_{3/2}^2 - \left(\frac{F^{T_b}}{2}\right)^2 \partial^2_{\tau_b} \ln\tilde{K}
\simeq \frac{5\left(c_s-\frac 13\right)}{\omega'_S}\, m_{3/2}M_{1/2}\sim \mc{O}\left(m_{3/2}^2\frac{\left(\ln\vo\right)^{3/2}}{\vo}\right)\,.
\label{localscalar}
\ee
The dominant contribution to this expression comes from the F-term of $T_b$.
More precisely, the leading term of $F^{T_b}$ in (\ref{FTb}) together with the leading term of $\tilde{K}$ in (\ref{newKtilde})
give a contribution which cancels against $m_{3/2}^2$ in (\ref{localscalar}) because of the underlying no-scale structure.
The first non-vanishing term in (\ref{localscalar}) originates from the leading term of $F^{T_b}$
together with the first subleading correction to $\tilde{K}$. On the other hand, the subleading correction to $F^{T_b}$ in (\ref{FTb}) yields a contribution
suppressed by $\epsilon_s$, and so turns out to be negligible.

Scalar masses are universal since they get generated by the F-term of $T_b$. Non-universal effects
can arise from $F^S$ and $F^{U_i}$ but they are volume suppressed since they would give contributions of order $m_{3/2}^2/\vo^2$. If $c_s>1/3,$ the scalar masses are non-tachyonic.

\medskip
\emph{Ultra-local limit}: An interesting feature of (\ref{localscalar}) is that it vanishes if one takes the ultra-local limit
$c_s= 1/3$.\footnote{We neglect potential higher order corrections at this stage.} In fact, there is a general argument \cite{Blumenhagen:2009gk} which guarantees the vanishing of $m_0^2$
at $\mc{O}(\vo^{-3})$. Using (\ref{ul}) (the defining property of ultra-local models)
in the general expression for the F-term contributions to scalar masses (\ref{genscalF}) we find
\be
\left.m_\alpha^2\right|_F = - \frac 13 V_{F,0} - F^I \overline{F}^{\overline{J}} \partial_I \partial_{\overline{J}} \ln h_\alpha(U,S)\,,
\label{ulsm2}
\ee
where we used the fact that $V_F= K_{I\overline{J}}F^I \overline{F}^{\overline{J}}- 3 m_{3/2}^2$.
Recalling that $V_0 = V_{F,0} + V_{D,0}$ and setting the cosmological constant to zero (in the dS constructions of Sec.~\ref{dsv} we showed how to cancel $V_0$ at $\mc{O}(\vo^{-3})$ but this can in principle be done at any order in the $\vo^{-1}$ expansion), $V_{F,0}$ can be traded for $V_{D,0}$,
and we so shall include it in our analysis of D-term contributions to scalar masses.

On the other hand, if the functions $h_\alpha(S,U)$ are not constants, there is a non-vanishing contribution
from the F-terms of the dilaton and the complex structure moduli at $\mc{O}(\vo^{-2})$. Using (\ref{ds}),
the $S$ and $U$-dependent contribution to scalar masses turns out to be
\be
\left.m_\alpha^2\right|_F =  - M_{1/2}^2 s^2 \left(\partial_s^2+\beta^{U_i}\partial_{u_i} \partial_s
+ \beta^{U_i}\beta^{\overline{U}_j}\partial_{u_i} \partial_{u_j}\right) \ln h_\alpha(U,S)\sim \mc{O}\left(M_{1/2}^2\right)\,,
\label{ulsm3}
\ee
where $M_{1/2}$ is the gaugino mass in (\ref{ggmm}). Note that this contribution is generically non-universal
and might also give rise to tachyonic scalars depending on the explicit functional dependence of the functions $h_\alpha(U,S)$.\\

\noindent \textbf{D-term contributions}
\medskip\\
Assuming a diagonal K\"ahler matter metric as in~\eqref{mm}, the general expression for the D-term contributions to scalar masses in gravity mediation is \cite{Dudas:2005vv}
\be
\left.m_\alpha^2\right|_D = \tilde{K}_\alpha^{-1} \sum_i g_i^2 D_i \partial^2_{\alpha \overline{\alpha}} D_i- V_{D,0}\,.
\label{genscalD}
\ee
Given that this result depends on the value of the D-term potential at the minimum, this contribution depends on the way to achieve a dS vacuum.
As explained in Sec.~\ref{shrink}, the VEV of the D-term potential associated to visible sector $U(1)$s is vanishing in the absence of tachyonic scalars.\footnote{Even in the presence of tachyonic scalars, the contribution to scalar masses from visible sector D-terms turns out to be a negligible effect since visible matter fields, $\tau_\SM$ and $b$ would still be stabilised at zero at leading order.}

\medskip
\emph{dS case 1}: In the dS case 1 of Sec.~\ref{matteruplift} the relevant D-term is the one associated with the anomalous $U(1)$ living on the big cycle.
As can be seen from (\ref{VD0ds1}), $V_{D,0}$ scales as $\vo^{-10/3}$ which is subdominant with respect to the first term in (\ref{genscalD}) that gives
\be
\left.m_0^2\right|_D = \frac{q_b}{2 f_\alpha(U,S)} D_{\dS_1} \partial_{\tau_b} \tilde{K}_{\alpha}= \frac{m_{3/2}^2}{3s}|\phi_\dS|^2
= \frac{6\epsilon_s}{\omega'_S}m_{3/2}M_{1/2}\sim  \mc{O}\left(m_{3/2}^2\frac{\sqrt{\ln\vo}}{\vo}\right)\,,
\label{sc1}
\ee
once we impose the condition (\ref{CCconddS1}) to have a vanishing cosmological constant at $\mc{O}(\vo^{-3})$.
In the local limit, this result is suppressed with respect to the F-term contribution (\ref{localscalar}) by a factor of $\epsilon_s$.
On the other hand, in the ultra-local limit, this D-term contribution dominates over the F-term one given in (\ref{ulsm3}) which scales as $m_{3/2}^2\epsilon^2$.
Hence it leads to universal and non-tachyonic scalar masses.

\medskip
\emph{dS case 2}: In the dS case 2 of Sec.~\ref{dilatonuplift} the relevant D-term is the one associated with the anomalous $U(1)$ which belongs to the hidden sector responsible for achieving a dS vacuum. In this case both terms in (\ref{genscalD}) have the same scaling since
\be
\left.m_0^2\right|_D = \frac{q_\dS \vo^{2/3}}{2 s f_\alpha(U,S)} D_{\dS_2} \partial_{\tau_\dS} \tilde{K}_{\alpha}-V_{D,0}
= \frac{c_\dS}{s} D_{\dS_2} q_\dS\frac{\tau_\dS}{\vo} -V_{D,0} = \left(2c_\dS-1\right)V_{D,0}\,.
\label{Vds22}
\ee
As can be seen from (\ref{VD0ds2}), $V_{D,0}$ scales as $\vo^{-4}$. Hence in the local limit the D-term contribution
is subleading with respect to the F-term one given in (\ref{localscalar}) which scales as $\vo^{-3}$.
In the ultra-local limit the F-term contribution to scalar masses is given by (\ref{ulsm2}).
Adding $-V_{F,0}/3= V_{D,0}/3$ to (\ref{Vds22}) we find that the total D-term contribution to scalar masses vanishes in the ultra-local limit once we impose $c_\dS=1/3$ as in (\ref{fdef}) since
\be
\left.m_0^2\right|_D =  2\left(c_\dS-\frac 13\right)V_{D,0}=0 \quad\text{for}\quad c_\dS=\frac 13\,.
\label{Vds2}
\ee
Hence scalar masses get generated by F-terms also in the ultra-local limit.
Their expression is given in (\ref{ulsm3}) and scales as $\vo^{-4}$.\\

\noindent \textbf{Summary}
\medskip\\
Let us summarise our results for soft scalar masses.
The expression for $m_0^2$ in the local limit does not depend on the way to obtain a dS vacuum since in each case
it is given by the F-term contribution (\ref{localscalar}) that scales as $\vo^{-3}$. Scalar masses are non-tachyonic if $c_s>1/3$ and universal.
On the other hand, the result for the ultra-local limit depends on the dS mechanism.
In the dS case 1 of Sec.~\ref{matteruplift}, scalar masses get generated by the D-term contribution (\ref{sc1}) which has again an overall $\vo^{-3}$ scaling but with an $\epsilon_s$ suppression with respect to the local case. Scalar masses turn out to be non-tachyonic and universal.
On the contrary, in the dS case 2 of Sec.~\ref{dilatonuplift}, the main contribution to scalar masses comes from F-terms and it is given by (\ref{ulsm3}) which scales as $\vo^{-4}$. This result could potentially lead to tachyonic and non-universal scalar masses depending on the exact functional dependence of the functions $h_\alpha(U,S)$.

\subsubsection{A-terms}

For the current discussion, we assume that the Yukawa couplings $Y_{\alpha \beta \gamma}$ receive no non-perturbative contributions from the K\"ahler moduli and are hence only functions of the complex structure moduli and the dilaton $Y_{\alpha \beta \gamma}=Y_{\alpha \beta \gamma}(U,S)$. The trilinear A-terms in gravity mediation receive both F- and D-term contributions. The D-term contributions turn out to be zero for vanishing VEVs of visible sector matter fields \cite{Dudas:2005vv}. On the other hand, the general formula for the F-term contribution is \cite{Brignole:1993dj}
\be
\label{tril}
A_{\alpha\beta\gamma} = F^I \partial_I
\left[K + \ln \left(\frac{Y_{\alpha \beta \gamma}(U,S)}{\tilde{K}_\alpha \tilde{K}_\beta \tilde{K}_\gamma}\right)\right]
= F^I \partial_I
\left[K -3\ln\tilde{K}+ \ln \left(\frac{Y_{\alpha \beta \gamma}(U,S)}{f_\alpha f_\beta f_\gamma}\right)\right],
\ee
where the holomorphic Yukawas $Y_{\alpha \beta \gamma}(U,S)$ do not depend on the K\"ahler moduli because of their axionic shift symmetry.
Let us present the expression for $A_{\alpha\beta\gamma}$ at leading order in $\vo^{-1}$ and $\epsilon_s$.

\medskip
\emph{Local limit}: In the local limit we find
\be
\label{tril1}
A_{\alpha\beta\gamma}= - \left[1 -s\beta^{U_i}\partial_{u_i} K_{\rm cs} - \frac{6}{\omega'_S}\left(c_s-\frac 13\right)
- s \partial_{s,u} \ln \left(\frac{Y_{\alpha \beta \gamma}}{f_\alpha f_\beta f_\gamma} \right) \right] M_{1/2}\sim\mc{O}\left(M_{1/2}\right)\,,
\ee
with $\partial_{s,u}\equiv \partial_s+\beta^{U_i}\partial_{u_i}$.
Note that there is a cancellation at $\mc{O}(\vo^{-1})$ between $K$ and $3\ln\tilde{K}$ in (\ref{tril}).
The dominant contributions to (\ref{tril1}) come from the F-terms of $T_b$, $S$ and $U$.

\medskip
\emph{Ultra-local limit}: In the ultra-local limit defined by (\ref{ul}), the contribution to $A_{\alpha\beta\gamma}$ from $F^{T_b}$ vanishes,
as can be seen at leading order in (\ref{tril1}) by setting $c_s=1/3$ and $f_{\alpha} = h_\alpha\,e^{K_{\rm cs}/3}(2s)^{-1/3}$.
In this limit, the general expression (\ref{tril}) simplifies to
\be
\label{trilul}
A_{\alpha\beta\gamma} = s \partial_{s,u}
\ln \left(\frac{Y_{\alpha \beta \gamma}(U,S)}{h_\alpha h_\beta h_\gamma}\right) M_{1/2}\sim\mc{O}\left(M_{1/2}\right)\,.
\ee

\subsubsection{$\hat{\mu}$ and $B\hat{\mu}$ terms}
\label{bmusection}

Let us discuss different effects that can contribute to the superpotential and K\"ahler potential Higgs bi-linear terms. Whether they are present or not is model dependent and a concrete realisation or combination of various mechanisms might not be possible. The following list should be understood as a list of possible effects that can lead to a non-vanishing $\mu$-term.
The canonically normalised $\hat{\mu}$ and $B\hat{\mu}$-terms receive contributions from both K\"ahler potential and superpotential effects.
Let us discuss these two different effects separately.\\

\noindent \textbf{K\"ahler potential contributions}\medskip\\
Non-zero $\hat{\mu}$ and $B \hat{\mu}$ get generated from a non-vanishing prefactor $Z$ in the matter K\"ahler potential (\ref{mm})~\cite{Kim:1983dt,Giudice:1988yz}. Their general expression in gravity mediation is \cite{Brignole:1993dj,Dudas:2005vv}
\be
\hat{\mu} = \left(m_{3/2} Z - \overline{F}^{\overline{I}} \partial_{\overline{I}} Z\right)
\left(\tilde{K}_{H_u} \tilde{K}_{H_d}\right)^{-1/2}\quad\text{and}\quad B\hat{\mu} = \left.B\hat{\mu}\right|_F + \left.B\hat{\mu}\right|_D\,,
\label{muK}
\ee
where
\begin{eqnarray}
\left.B\hat{\mu}\right|_F &=& \left\{2 m_{3/2}^2 Z - m_{3/2} \overline{F}^{\overline{I}} \partial_{\overline{I}} Z + m_{3/2} F^I \left[\partial_I Z - Z \partial_I \ln\left(\tilde{K}_{H_u} \tilde{K}_{H_d}\right)\right]\right. \nonumber \\
&& \left. - F^I \overline{F}^{\overline{J}} \left[\partial_I \partial_{\overline{J}} Z - \partial_I Z \partial_{\overline{J}} \ln\left(\tilde{K}_{H_u} \tilde{K}_{H_d}\right)\right] \right\}\left(\tilde{K}_{H_u} \tilde{K}_{H_d}\right)^{-1/2} \,, \label{BmuKF} \\
\left.B\hat{\mu}\right|_D &=& \left(\tilde{K}_{H_u} \tilde{K}_{H_d}\right)^{-1/2}\left(\sum_i g_i^2 D_i \partial_{H_u} \partial_{H_d} D_i - V_{D,0} Z\right)\,.
\label{BmuKD}
\end{eqnarray}
Motivated by the fact that we are at the singular regime, we take $Z$ of the same form as the matter metric (\ref{mattermetric})
with $f_\alpha(U,S)$ replaced by a different unknown function of $S$ and $U$ which we denote $\gamma(U,S)$.
We stress that $Z=\gamma(U,S)\tilde{K}$ is just the simplest ansatz for $Z$ given our present knowledge
but its form could in general be different from $\tilde{K}_\alpha$.\footnote{However in models with a shift-symmetric Higgs sector $f_{H_u}=f_{H_d}=\gamma$ \cite{LopesCardoso:1994is,Antoniadis:1994hg,Brignole:1995fb,Brignole:1996xb,Hebecker:2012qp}.}

Let us compute the leading expressions (in an expansion in $\vo^{-1}$ and $\epsilon_s$) for both $\hat\mu$ and $B\hat\mu$
in the local and ultra-local limit.

\medskip
\emph{Local limit}: In the local limit we find
\be
\hat\mu  =  \frac{\gamma}{\sqrt{f_{H_u } f_{H_d}}} \left[\frac{6}{3 \omega'_S} \left(c_s-\frac 13\right)
- s \partial_{s,u} \ln\gamma \right] M_{1/2}\sim\mc{O}\left(M_{1/2}\right)\,,
\label{mu11}
\ee
where again there is a cancellation at $\mc{O}(\vo^{-1})$ between the term proportional to $m_{3/2}$ in (\ref{muK})
and the leading order contribution from $\overline{F}^{\overline{T}_b}\partial_{\overline{T}_b}Z$.
The dominant contributions to (\ref{mu11}) come from the F-terms of $T_b$, $S$ and $U$.
On the other hand, the $B\hat\mu$-term behaves as the soft scalar masses since both F- and D-term contributions can be rewritten as
\be
\left.B\hat{\mu}\right|_{F,D} = \frac{\gamma}{\sqrt{f_{H_u} f_{H_d}}}\, \left.m_0^2\right|_{F,D}\,.
\label{bmu11}
\ee
Recalling our results for $m_0^2$, we realise that in the local limit the leading contribution to $B\hat\mu$ comes from F-terms
and scales as $\left.m_0^2\right|_F$ in (\ref{localscalar}). Hence the final result for $B\hat\mu$ is
\be
B\hat{\mu} = \frac{\gamma}{\sqrt{f_{H_u} f_{H_d}}}\, \frac{5\left(c_s-\frac 13\right)}{\omega'_S} m_{3/2}M_{1/2}\sim \mc{O}\left(m_{3/2}^2\frac{\left(\ln\vo\right)^{3/2}}{\vo}\right)\,.
\label{bmu11local}
\ee

\medskip
\emph{Ultra-local limit}: Similarly to the ultra-local limit for $\tilde{K}_\alpha$ defined by (\ref{ul}), we can define also an ultra-local limit
for $Z=\gamma(U,S)\tilde{K}$ as $Z\equiv z(U,S)\,e^{K/3}$ which implies
\be
\label{fdef2}
\gamma(U,S) = \frac{z(U,S)\,e^{K_{\rm cs}/3}}{(2s)^{1/3}}\qquad \text{and}\qquad c_s=c_\dS=\frac 13\,.
\ee
In this limit the F-term of $T_b$ does not contribute to $\hat\mu$ whose expression simplifies to
\be
\label{mu2}
\hat\mu  =  - \frac{z\,s\partial_{s,u} \ln\gamma}{\sqrt{h_{H_u } h_{H_d}}}\,M_{1/2}\sim\mc{O}\left(M_{1/2}\right)\, .
\ee
In this ultra-local case the expression (\ref{BmuKF}) for $\left.B\hat\mu\right|_F$ gives
\be
\left.B\hat\mu\right|_F = \frac{z}{\sqrt{h_{H_u } h_{H_d}}}\left[\sigma(U,S) \,M_{1/2}^2-\frac 13 V_{F,0}\right]\,,
\label{Bmu2}
\ee
where $\sigma(U,S)$ is a complicated $\mc{O}(1)$ function of $S$ and $U$ which looks like
\begin{eqnarray}
\sigma(U,S) &=& \frac 19 \left(1-s\beta^{U_i}\partial_{u_i}K_{\rm cs} \right)
\left[1 - 3 s \partial_{s,u} \ln \left(h_{H_u} h_{H_d}\right)\right] \nonumber \\
&+& s \partial_{s,u} \ln \left(h_{H_u} h_{H_d}\right) s \partial_{s,u}\ln z- s^2\left[\partial_s \ln z \,\partial_{s,u} \ln \partial_s z
+\beta^{U_i} \partial_{u_i} \ln z \,\partial_{s,u} \ln \partial_{u_i} z\right]\,. \nonumber
\end{eqnarray}
Recalling that $V_0 = V_{F,0} + V_{D,0}=0$, $V_{F,0}$ can be traded for $V_{D,0}$,
and so we shall include it in our analysis of D-term contributions to $B\hat\mu$.
\begin{enumerate}
\item In the dS case 1 of Sec.~\ref{matteruplift}, the D-term generated $B\hat\mu$ is
\be
\left.B\hat\mu\right|_D = \frac{z}{\sqrt{h_{H_u} h_{H_d}}}\,\left.m_0^2\right|_D
= \frac{z}{\sqrt{h_{H_u} h_{H_d}}}\,\frac{6\epsilon_s}{\omega'_S}m_{3/2}M_{1/2}\sim  \mc{O}\left(m_{3/2}^2\frac{\sqrt{\ln\vo}}{\vo}\right)\,,
\label{bmu2dS1}
\ee
where we used the result in (\ref{sc1}). This term dominates over the F-term contribution given in (\ref{Bmu2}).

\item In the dS case 2 of Sec.~\ref{dilatonuplift}, the D-term generated $B\hat\mu$ is vanishing since
\be
\left.B\hat\mu\right|_D = \frac{z}{\sqrt{h_{H_u} h_{H_d}}}\,\left.m_0^2\right|_D
= \frac{z}{\sqrt{h_{H_u} h_{H_d}}}\,2\left(c_\dS-\frac 13\right)V_{D,0}=0 \quad\text{for}\quad c_\dS=\frac 13\,,
\label{bmu2dS2}
\ee
where we used the result in (\ref{Vds2}). Thus in this case $B\hat\mu$ is generated purely by F-terms and it is given by (\ref{Bmu2})
without the term proportional to $V_{F,0}$ that we included in the D-term contribution. Hence the final result for $B\hat\mu$ is
\be
B\hat\mu = \frac{z}{\sqrt{h_{H_u} h_{H_d}}}\,\sigma(U,S) \,M_{1/2}^2\sim\mc{O}\left(M_{1/2}^2\right)\,.
\label{Bmu22}
\ee
\end{enumerate}
\newpage
\noindent \textbf{Superpotential contributions}\medskip\\
Let us discuss the contributions to $\hat\mu$ and $B\hat\mu$ from $\mu\neq 0$ in $W_{\rm matter}$ given by (\ref{Wmatter}).
Their general expression in gravity mediation reads \cite{Brignole:1993dj}
\begin{eqnarray}
\label{muW}
\hat{\mu} &=& \mu\,e^{K/2} \left(\tilde{K}_{H_u} \tilde{K}_{H_d}\right)^{-1/2}\,, \\
\label{BmuW}
B\hat{\mu} &=& \mu\,e^{K/2} \left[F^I \left(K_I + \partial_I \ln\mu - \partial_I \ln\left(\tilde{K}_{H_u} \tilde{K}_{H_d}\right)\right) - m_{3/2}\right]\left(\tilde{K}_{H_u} \tilde{K}_{H_d}\right)^{-1/2}\,.
\end{eqnarray}

\medskip
\emph{Non-perturbative effects}: Non-perturbative effects can generate in the low-energy action an effective $\mu$-term of the form (up to prefactors)
\be
W\supset e^{-a T} H_u H_d\quad\Rightarrow\quad \mu_{\rm eff} = e^{-a T}\,,
\label{Wnpmu}
\ee
if the cycle $\tau={\rm Re}(T)$ is in the geometric regime \cite{Ibanez:2007tu} or
\be
W\supset e^{-b \left(S+\kappa T\right)} H_u H_d\quad\Rightarrow\quad \mu_{\rm eff} = e^{-b\left(S+\kappa T\right)}\,,
\label{Wnpmu2}
\ee
if the cycle $\tau={\rm Re}(T)$ is in the singular regime, i.e. $\tau\to 0$ \cite{Berenstein:2005xa}. Note that there are two distinct classes of non-perturbative contributions leading to the above EFT coupling: if the Higgs bi-linear is forbidden by anomalous $U(1)$ symmetries, charged instanton contributions for instance via ED3 can realise such a coupling~\cite{Ibanez:2007tu,Berenstein:2005xa}. Alternatively, if the Higgs-bilinear is forbidden by an approximate global symmetry of the local model, this global symmetry is broken by compactification effects. For the latter case, ref.~\cite{Berg:2012aq} studied the topological conditions under which non-perturbative effects of the form (\ref{Wnpmu}) and (\ref{Wnpmu2})
contribute to the effective action. If $T$ is a bulk cycle, the coupling (\ref{Wnpmu}) is always generated but in our case it would be negligible since this effect would be proportional to $e^{-\vo^{2/3}}$. On the other hand, if $T$ is the blow-up of a local singularity, the couplings (\ref{Wnpmu}) and (\ref{Wnpmu2})
get generated only if this divisor shares a homologous two-cycle with the blow-up mode $T_\SM$ of the MSSM singularity.
This condition is not satisfied if either $T$ or $T_\SM$ is a very simple divisor like a dP$_0$ which has been used in the explicit global models of \cite{Cicoli:2012vw} and \cite{Cicoli:2013cha}.

If in both cases the appropriate conditions are satisfied, both (\ref{Wnpmu}) and (\ref{Wnpmu2}) would lead to a non-vanishing contribution which can be parameterised as follows
\be
\hat{\mu} \simeq \frac{c_\mu(U,S)}{\vo^{n+\frac 13}}
\qquad\text{and}\qquad
B \hat{\mu} \simeq \frac{c_B(U,S)}{\vo^{n+ \frac 43}}\,,
\label{mumag}
\ee
where in (\ref{Wnpmu}) we have set $T=T_s$ and $a=n a_s$ with $n>0$, while in (\ref{Wnpmu2}) we have parameterised $b=n a_s$
recalling that $s\simeq \tau_s$ from (\ref{tausms}). $c_{\mu}$ and $c_B$ are constants which absorb the dependence on the prefactor of the instanton contribution, the complex structure moduli and the dilaton. Note that for different values of $n$ non-perturbative effects could generate $\hat{\mu}$ and $B \hat{\mu}$ in the complete range interesting for MSSM phenomenology regardless of the size of the other soft-terms.
However these effects can be in competition with K\"ahler potential contributions for $n \geq 5/3$.\\

\emph{Background fluxes}: Primitive $(1,2)$ IASD fluxes
can generate $\hat\mu$ for D3-branes at singularities \cite{Grana:2002nq,Camara:2003ku,Grana:2003ek}.
However, given that their contribution is proportional to the F-terms of the complex structure moduli,
this effect has already been included in the contributions from the K\"ahler potential. In other words, direct computations of soft terms
by reducing the D3-brane action in a fluxed background show that $\mu=0$ \cite{Grana:2003ek}.\\

\noindent \textbf{Anomalous $U(1)$ symmetries}\medskip\\
A term proportional to $H_u H_d$ in $K$ or $W$ could be forbidden in the presence of an anomalous $U(1)$ symmetry.
In this case, the only way to generate a Higgs bilinear would be to multiply this term by an operator involving a $U(1)$-charged field
which makes the whole contribution gauge invariant. As already discussed above, the only closed string moduli that can lead to such an effect are K\"ahler moduli.

Alternatively, the $U(1)$-charged field could be an open string mode $\Phi$ appearing in $K$ and $W$
in a gauge invariant combination of the form ($\Lambda$ denotes the appropriate moduli-dependent cut-off)
\be
K\supset \left(\frac{\Phi}{\Lambda}\right)^n H_u H_d\,,\quad \quad W\supset\frac{\Phi^n}{\Lambda^{n-1}} H_u H_d\,.
\label{KWphi}
\ee
Thus the field $\Phi$ has to be an MSSM singlet since a Higgs bilinear gets generated only when $\Phi$ develops a non-zero VEV breaking the $U(1)$ symmetry.
However, as can be seen from eqs. (\ref{dtermpot}) and (\ref{xiSM}), D-term stabilisation fixes the VEV of $\Phi$
in terms of $\tau_\SM$: $|\Phi|^2 \propto \tau_\SM/\vo$,
and so the couplings in (\ref{KWphi}) would give rise to effective $\mu$ and $Z$-terms which depend only on closed string moduli
\be
Z_{\rm eff}\propto \frac{1}{\Lambda^n}\left(\frac{\tau_\SM}{\vo}\right)^{n/2}\,,
\quad \quad\mu_{\rm eff}\propto \frac{1}{\Lambda^{n-1}}\left(\frac{\tau_\SM}{\vo}\right)^{n/2}\,.
\label{Eff}
\ee
Once the cut-off $\Lambda$ is explicitly written in terms of $T$-moduli, one could plug (\ref{Eff})
into the standard supergravity formulae to work out the final contribution to $\hat\mu$ and $B\hat\mu$.
The result will depend on the VEV and the F-term of $T_\SM$. As discussed in \cite{Cicoli:2013cha},
$\Phi$ needs to receive tachyonic contribution from soft terms in order for $T_\SM$ to develop a non-zero VEV.
If this condition is satisfied, $\tau_\SM \sim \vo^{-1}$ implying $F^{T_\SM}\sim \vo^{-2}$ for the local case
and $\tau_\SM \sim \vo^{-3}$ implying $F^{T_\SM}\sim \vo^{-4}$ for the ultra-local case.
This effect corresponds to switching on an FI-term, and so to breaking the anomalous $U(1)$ by moving slightly away from the singularity.
However in both cases the VEV of $\tau_\SM$ is smaller than unity, and so we are still consistently in the singular regime.

Given that all these results are clearly model-dependent and require physics beyond the MSSM,
at this stage we do not pursue these options in more detail and leave them for future work.
Let us just mention that the only case where the effective $\mu$-term in (\ref{Eff}) does not
depend on $\Lambda$ is for $n=1$. In this situation $\hat\mu$ would scale as $\vo^{-4/3}$ in the local case and as $\vo^{-7/3}$ in the ultra-local case.
If instead $\Phi$ does not receive tachyonic contributions from soft terms,
another option would be to consider models where $\Phi$ develops a non-zero VEV because of radiative effects.

\subsection{Summary of soft-terms}

Let us summarise our results for the soft-terms in the two cases to obtain dS vacua (see also Table \ref{Table1}).
Given that in each case the gaugino masses turn out to have the same value,
we will use $M_{1/2}$ as a useful parameter which can be rewritten as
\be
M_{1/2} = c_{1/2} m_{3/2} \frac{m_{3/2}}{M_P}\left[\ln\left(\frac{M_P}{m_{3/2}}\right)\right]^{3/2}\ll m_{3/2}\, ,
\ee
where $c_{1/2}$ is a flux-dependent tunable coefficient.
We will state our results for the model-independent case where $\hat\mu$ and $B\hat\mu$ are generated
from moduli induced K\"ahler potential contributions. If these contributions are absent (for example if these terms are forbidden by anomalous $U(1)$ symmetries), then $\hat\mu$ and $B\hat\mu$ can take different values because of model-dependent contributions from either $K$ or $W$
as discussed in Sec.~\ref{bmusection}. Let us discuss the local and ultra-local limits separately.

\medskip
\emph{Local limit}: In the local limit, the soft-terms turn out to be the same in both dS mechanisms
(all the $c$'s are flux-dependent parameters)
\be
m_0^2 = c_0\,m_{3/2} M_{1/2}\,,\quad A_{\alpha\beta\gamma} = (c_A)_{\alpha\beta\gamma}\,M_{1/2}\,,
\quad\hat\mu= c_\mu\,M_{1/2}\,,\quad B\hat\mu = c_B m_0^2\,.
\ee

\medskip
\emph{Ultra-local limit}: In the ultra-local limit, the soft-terms take different forms in the two dS cases
(again all the $c$'s are flux-dependent coefficients which are distinct in different scenarios)
\begin{enumerate}
\item dS vacua from hidden matter fields
\be
m_0^2 = c_{0}\,\frac{m_{3/2}M_{1/2}}{\ln\left(M_P/m_{3/2}\right)}\,,\quad A_{\alpha\beta\gamma} = (c_A)_{\alpha\beta\gamma}\,M_{1/2}\,,
\quad\hat\mu= c_\mu\,M_{1/2}\,,\quad B\hat\mu = c_{B} m_0^2\,;
\ee
\item dS vacua from non-perturbative effects at singularities
\be
m_\alpha = (c_{0})_\alpha\,M_{1/2}\,,\quad A_{\alpha\beta\gamma} = (c_A)_{\alpha\beta\gamma}\,M_{1/2}\,,
\quad\hat\mu= c_\mu\,M_{1/2}\,,\quad B\hat\mu = c_{B} M_{1/2}^2\,.
\ee
\end{enumerate}
Clearly, the local limit and the dS case 1 for the ultra-local limit correspond to typical (mini) split SUSY scenarios whereas
the dS case 2 for the ultra-local limit reproduces a standard MSSM picture with universal gaugino masses and soft masses all of the same order. If the flux dependent coefficients for the scalar masses are universal $(c_0)_\alpha=c_0,$ a standard CMSSM scenario emerges. Non-universalities in the flux dependent coefficients can lead to interesting soft-term patterns such as in NUHM or natural SUSY scenarios. We will study in detail the LHC phenomenology
of these different scenarios in a subsequent paper~\cite{us}.

\begin{table}
\begin{center}
{\tabulinesep=1.4mm
   \begin{tabu}{|c|c|c|c|}
\hline
Soft term &  Local Models & Ultra Local dS$_1$ & Ultra Local dS$_2$ \\ \hline \hline
$M_{1/2}$ & \multicolumn{3}{c|} { $c_{1/2}\,m_{3/2}\,\frac{m_{3/2}}{M_P}\, \left[ \ln\left(\frac{M_P}{m_{3/2}}\right)\right]^{3/2}$}\\ \hline
$m_{\alpha}^2$ & $c_0\, m_{3/2} M_{1/2}$ &  $c_0\, \frac{m_{3/2}M_{1/2}}{\ln (M_P/m_{3/2})}$ & $(c_0)_\alpha\, M_{1/2}^2$ \\ \hline
$A_{\alpha\beta\gamma}$ & \multicolumn{3}{c|}{$(c_A)_{\alpha\beta\gamma}\, M_{1/2} $}
\\ \hline
$\hat\mu$ & \multicolumn{3}{c|}{$\begin{array}{c}c_\mu\, M_{1/2}\, \qquad\qquad\qquad \text{ (contribution from $K$)} \\
c_\mu M_P\left[\frac{m_{3/2}}{M_P}\right]^{n+1/3}\, \qquad \text{(contribution from $W$)}
\end{array}$}
\\ \hline
$B\hat\mu$ & \multicolumn{3}{c|}{$\begin{array}{c}c_B\, m_{0}^2\, \qquad\qquad\qquad  \text{ (contribution from $K$)} \\
c_B m_{3/2}\left[\frac{m_{3/2}}{M_P}\right]^{n+1/3} \quad \text{(contribution from $W$)}
\end{array}$}
\\
\hline
\end{tabu}}
\caption{Summary of soft-terms for different sequestered scenarios for the two dS mechanisms discussed in the text:
hidden sector matter (dS$_1$) and non-perturbative effects at singularities~(dS$_2$). All soft terms are hierarchically smaller than $m_{3/2}$.
Gaugino masses, A-terms and the $\hat{\mu}$-term take the same value in each case whereas scalar masses and hence the $B\hat{\mu}$-term are model-dependent. The coefficients $c$ are flux dependent and can generically take different values in each scenario presented here while $n$ is a positive model-dependent parameter. They can be tuned to compare with data. Local and ultra-local 1 cases give a split SUSY spectrum while ultra local 2 implies a standard MSSM spectrum with soft-masses of the same order and possible small non-universalities due to the flux dependent parameters $c$.}
\label{Table1}
\end{center}
\end{table}

For illustrative purposes, we just mention here two simple benchmark models for the dS case 2 in the ultra-local limit.
Setting all the $\beta$'s to zero, we find

\medskip
\emph{Benchmark model 1: $h_\alpha=z=1$}
\be
m_\alpha \simeq 0\,\,\,\forall\alpha\,,\quad A_{\alpha\beta\gamma} = (c_A)_{\alpha\beta\gamma}\,M_{1/2}\,,
\quad\hat\mu= \frac{M_{1/2}}{3} \,,\quad B\hat\mu = \hat\mu^2\,,
\ee
where $(c_A)_{\alpha\beta\gamma}=s\partial_s \ln Y_{\alpha\beta\gamma}$.
This reproduces a typical gaugino mediation scenario \cite{Schmaltz:2000gy,Yanagida:2013ah}.

\medskip
\emph{Benchmark model 2: $f_\alpha=\gamma=1$}
\be
m_\alpha = m_0 = \frac{M_{1/2}}{\sqrt{3}}\,\,\,\forall\alpha\,,\quad A = -\sqrt{3}\,m_0\,,
\quad\hat\mu \simeq \frac{m_0}{\ln\left(M_P/m_{3/2}\right)}\,,\quad B\hat\mu = m_0^2\,,
\ee
if the holomorphic Yukawas do not depend on $S$.
This leads to a typical natural SUSY scenario for example
if we allow $m_{H_u}$ to be slightly larger than $m_0$ together with a light third generation~\cite{Baer:2012uy}.
This can be done by considering the more general case with non-zero
$\beta$'s and allowing for a $U$-dependence in $f_\alpha$. The $\ln\left(M_P/m_{3/2}\right)$ suppression of $\hat\mu$ with respect to $m_0$ comes from subleading contributions to $\hat\mu$ from $F^{T_s}$.

\section{Conclusions}
\label{sec:conclusions}

In this paper we have analysed soft-terms for LVS sequestered models with dS moduli stabilisation.
These models are particularly attractive for phenomenological reasons: the string scale is
around the GUT scale, soft masses are at the TeV scale and the lightest modulus is much heavier than the bound imposed by the cosmological moduli problem.
The volume of the compactification is of order $\vo\sim 10^7$ in string units and the visible sector can be localised on D3-branes at a singularity.

The pattern of soft terms for these models has been previously studied in \cite{Blumenhagen:2009gk}. In this paper we have deepened the analysis of \cite{Blumenhagen:2009gk} by studying the effect on soft terms of the sector responsible to realise a dS vacuum, and by classifying in a systematic way any possible correction to the leading no-scale structure of soft terms. In particular, given that soft terms depend on the moduli-dependence of the K\"ahler metric for matter fields $\tilde{K}_\alpha$, we defined two possible limits for $\tilde{K}_\alpha$: \emph{Local scenarios} where $\tilde{K}_\alpha$ is such that the visible sector Yukawa couplings $Y_{\alpha\beta\gamma}$ do not depend on $\vo$ only at leading order in an inverse volume expansion, and \emph{Ultra-local scenarios} where $Y_{\alpha\beta\gamma}$ are exactly independent on $\vo$ at all orders.\footnote{Evidence in favour of ultralocality has been obtained from explicit string computations in toroidal orbifolds \cite{Conlon:2011jq}.
The case of realistic CY compactifications remains to be explored.} Moreover, due to the present lack of explicit string computations of $\tilde{K}_\alpha$, we parameterised its dependence on the dilaton and complex structure moduli as an unknown function $f_\alpha(U,S)$.

The computation of soft terms has produced a wide range of phenomenological possibilities depending on the exact moduli-dependence of the matter K\"ahler metric and the way to achieve a dS vacuum. We considered two dS realisations based on supersymmetric effective actions: \emph{dS case 1} where hidden sector matter fields living on a bulk cycle develop non-vanishing F-terms because of D-term fixing, and \emph{dS case 2} where the blow-up mode of a singularity different from the visible sector one develops non-zero F-terms due to non-perturbative effects.
Broadly speaking, we found two classes of models:
\begin{enumerate}
\item \emph{Split SUSY}: Local models and ultra-local models in the dS case 1 yield gaugino masses and A-terms which are suppressed with respect to scalar masses: $M_{1/2} \sim m_{3/2}\epsilon\ll m_0 \sim m_{3/2} \sqrt{\epsilon}\ll m_{3/2}$
    for $\epsilon \sim m_{3/2}/M_P\ll 1$. For volumes of order of $10^7$ in string units these models provide a version of the split SUSY scenario with a `largish' splitting between gauginos and scalars (according to current experimental bounds). Non-universalities are present but suppressed by inverse powers of the volume.

\item \emph{Standard MSSM}: For ultra-local models in the dS case 2, all soft-terms are of the same order: $M_{1/2} \sim m_0 \sim m_{3/2}\epsilon\ll m_{3/2}$.
Therefore these models include the CMSSM parameter space and its possible generalisations since each soft-term comes along with a tunable flux-dependent coefficient. Moreover, depending on the exact functional dependence of the K\"ahler metric for matter fields, these models can also feature non-universalities which are constrained by the experimental bounds on flavour changing neutral currents.
\end{enumerate}

Let us stress again that the exact numerical coefficients of the soft terms are functions of the dilaton and complex structure moduli which are fixed in terms of flux quanta. Hence soft terms vary as one scans through the string landscape. This crucial property of our scenarios gives supersymmetric models the freedom to perform any tuning which is needed for phenomenological reasons. In particular, it is low energy SUSY that addresses the hierarchy problem by stabilising the Higgs mass at the weak scale, while scanning through the landscape provides small variations in the size of soft terms as necessary to reproduce all the detailed features of experimental data. This tuning at low energies can be viewed as a choice of parameters in the high scale theory. There is a large freedom of choice in the high scale theory which is however not arbitrary since this freedom is provided by the theory itself (by having a computable landscape of vacua).

Note that in the ultra-local case the two ways to achieve a dS vacuum give rise to a different pattern of soft terms. This can intuitively be understood as follows: the depth of the LVS AdS vacuum is of order $m_{3/2}^2\epsilon$, and so any extra term which yields a dS solution has to be of this order of magnitude. In turn, if the field $\phi$ responsible for dS uplifting is not decoupled from the visible sector, scalar masses of order $m_{3/2} \sqrt{\epsilon}$ are expected to arise because of this new contribution to the scalar potential. This is actually what happens in the dS case 1 since $\phi$ lives on a bulk cycle, and so it is not decoupled from the visible sector. On the other hand, in the dS case 2 $\phi$ lives on a singularity which is geometrically separated from the one supporting the visible sector. This gives rise to an effective decoupling between $\phi$ and the visible sector, resulting in suppressed scalar masses.

We would also like to emphasise that our analysis for the dS case 1, together with~\cite{Cicoli:2012vw}
(which provided visible sector models embedded in moduli stabilised compact CYs),
provides a very comprehensive study of SUSY breaking in string theory.

The soft terms which are more complicated to estimate are the $\hat{\mu}$ and $B\hat{\mu}$-terms since they receive contributions from both the K\"ahler potential and the superpotential. Moreover, these contributions could generically be forbidden in models with branes at singularities because of the presence of anomalous $U(1)$ symmetries. In this case, effective $\hat{\mu}$ and $B\hat{\mu}$-terms could still be generated due to non-perturbative corrections (e.g. D-brane instantons) or matter fields which develop non-vanishing VEVs. However in this last case, besides the need to go beyond the MSSM by including additional matter fields, any prediction for $\hat{\mu}$ and $B\hat{\mu}$-terms would necessarily be model-dependent.

Overall, we are living exciting times with plenty of feedback from experiments. A detailed study of the phenomenology of the general sequestered scenarios mentioned above will be presented in a follow-up article \cite{us}.

\section*{Acknowledgments}

We thank Joe Conlon, Emilian Dudas, Bhaskar Dutta, Luis Ib\'a\~nez, Roberto Valandro, Giovanni Villadoro and Martin Winkler for useful discussions.
MC, SK, AM and FM  would like to thank the ICTP for hospitality.
The work of SK is supported by the DFG under TR33 ``The Dark Universe''. The work of AM is supported in part by a Ramanujan fellowship.

\appendix

\section{Subleading corrections to F-terms}
\label{App}

In this appendix we first describe the shift of the LVS minimum after including an extra term responsible to achieve a dS vacuum,
and then provide subleading corrections to F-terms. As described in Sec.~\ref{dsv}, the mechanism which realises a dS vacuum gives rise effectively to an extra term of the form
\be
V_\dS=\frac{r}{\vo^m}\qquad\text{with}\quad r>0\quad\text{and}\quad m<3\,.
\ee
We are interested in minimising the combined system
\be
V=V_{\mc{O}(\vo^{-3})}+V_\dS\,,
\ee
with the additional constraint of vanishing vacuum energy. This constraint relates the coefficient $r$ with the tunable flux parameters in the LVS potential such as $W_0$ or $g_s$. A concrete dS scenario, such as the ones in Sec.~\ref{dsv}, typically fixes $r$ by construction with only moderate tuning. However the real tuning can be achieved by simply tuning the flux superpotential and the string coupling in agreement with the flux landscape.

The expressions for the moduli VEVs are largely independent on the way to get dS vacua.
In fact, the relation (\ref{eataus}) is generic whereas the expression (\ref{tausms}) for the VEV of $\tau_s$
depends on the way to get a dS vacuum. The exact minimum for $\tau_s$ is given by
\be
\tau_s^{3/2}=\frac{\hat{\xi}}{2}\, \frac{(1-\epsilon_s)^2}{(1-4\epsilon_s)}\, \frac{1}{1+\frac{2m}{m-3} \epsilon_s}
=\frac{\hat{\xi}}{2}\left[1+f_\dS(\epsilon_s)\right]\,,
\label{eq:shiftds}
\ee
and so the function $f_\dS$ is $f_\dS= 18 \epsilon_s+297 \epsilon_s^2$ in the case of dS vacua from hidden matter fields ($m=8/3$),
while $f_\dS=3 \epsilon_s+12 \epsilon_s^2$ for the case of non-perturbative effects at singularities ($m=1$).
Note that as a consequence of the shift in $\tau_s$, also the overall volume in (\ref{eataus}) is shifted and, as the shift is in the exponential,
this shift can be parametrically large.

Equipped with the minimum, we can evaluate the F-terms. To simplify the notation we factor out an overall factor of the gravitino mass $m_{3/2}$ which is given by (\ref{expandm32}). The F-terms turn out to have the following expressions:
\bea
\frac{F^{T_b}}{\tau_b}&=&-2 m_{3/2}\left[1+\frac{9 \hat{\xi} \epsilon_s}{4 \vo} \frac{m-1}{m-3+2m\epsilon_s} +\mc{O}\left(\frac{1}{\vo^2}\right)\right]\,, \\
\frac{F^{T_s}}{\tau_s}&=& -2 m_{3/2}\left[\frac{3\epsilon_s}{(1-\epsilon_s)}-\frac{\hat{\xi}}{2\vo}
\left(1-\frac{9\epsilon_s}{2}\frac{m-1}{m-3}+\mc{O}\left(\epsilon_s^2\right)\right)\right]\,, \\
\frac{F^S}{s}&=& \frac{3 \omega'_S }{8 a_s^{3/2}}  \frac{m_{3/2}}{\vo\epsilon_s^{3/2}}\left[1+\mc{O}\left(\epsilon_s\right)\right]\,, \\
F^U&=& - \frac{K^{U_i \overline{U}_{\overline{j}}}}{2 s^2}\frac{\omega_{\overline{U}_{\overline{j}}}}{\omega'_S}F^S\equiv \beta^{U_i}\,F^S\,, \\
F^{\phi_\dS} &=& \phi_\dS m_{3/2}\left[1+ \mc{O}\left(\frac{1}{\vo}\right)\right] \,, \\
F^{T_{\dS}}&=& \frac{3}{4\sqrt{2}a_s^{3/4}}\frac{m_{3/2}}{\epsilon_s^{1/4}}\left[1+\mc{O}\left(\epsilon_s\right)\right]\,.
\eea

\section{Possible sources of desequestering}
\label{sec:deseq}

There is a general belief that in any supergravity theory once SUSY is broken all sparticles should get a mass at least of the order of the scale determined by the split in the gravity multiplet. In particular, all soft masses are expected to be of order the gravitino mass. Furthermore, if for some reason some of the sparticle masses are found to be smaller than $m_{3/2}$ at tree level, since SUSY no longer protects these masses against quantum corrections, they should be lifted to a loop factor times $m_{3/2}$. So soft masses are expected to be at most one order of magnitude lighter than the gravitino mass but not much smaller.\footnote{This separation between $m_{3/2}$ and soft masses occurs for example in the case of mirage (mixed moduli and anomaly) mediation~\cite{LoaizaBrito:2005fa}.}
Effects which tend to push the soft masses to the scale of the gravitino mass are referred to as sources of desequestering.
In this appendix we will argue that our models can be stable against desequestering effects.

\subsection{Loop corrections}

For sequestered string scenarios, it is natural to expect that loop corrections bring soft masses to a magnitude of order a loop factor times $m_{3/2}$. However there can be exceptions since the couplings can be Planck suppressed. A detailed calculation of loop corrections to the mass of bulk scalars like the volume modulus (its tree level mass $m_\vo \sim m_{3/2}/\vo^{1/2}$ is hierarchically smaller than $m_{3/2}$) was presented in~\cite{Burgess:2010sy}.

The size of loop corrections can be estimated by realising that, if SUSY is broken,
loop corrections to the mass should be given by the heaviest particles circulating in the loop
(or the cut-off scale) which is the Kaluza-Klein scale $M_\KK\sim M_P/\vo^{2/3}$.
In the absence of SUSY there is a need of a SUSY breaking insertion (a spurion field representing the relevant F-term)
in the loop and the correction to the mass is at most
\be
\delta m=\alpha_\lp \frac{M_\KK m_{3/2}}{M_P}\sim \alpha_\lp\frac{W_0}{\vo^{5/3}}\ll \alpha_\lp m_{3/2}\,,
\ee
with $\alpha_\lp\sim g^2/(16\pi^2)$ a loop factor. Note that the ratio $\delta m/m\sim \alpha_\lp \vo^{-1/6}$ is very small
and therefore the volume modulus mass is stable against loop corrections.

For matter fields located at the SM brane, loop corrections should be even further suppressed.
The effective field theory on the brane is supersymmetric and feels the effects of SUSY breaking in the bulk only via Planck suppressed couplings.
Therefore masses as small as $M_{\rm soft}\sim W_0/\vo^2$ are still stable under standard loop corrections
(since volume suppressed brane-bulk couplings imply $\delta M_{\rm soft}\ll \delta m$).

Over the years explicit calculations have been performed estimating loop corrections to soft masses in no-scale
and general gravity mediated models. See for example \cite{Ellis:1986nr,Antoniadis:1997ic}
in which loop corrections to scalar and gaugino masses were estimated in supergravity and M-theory frameworks with results of order
$\delta m \sim \alpha_\lp m_{3/2}^2/M_P\sim \alpha_\lp M_P/\vo^2$. More recently, explicit calculations for gravitino loop contributions to gaugino masses was performed in \cite{Lee:2013aia}. The diagrams are quadratically divergent and proportional to the gravitino mass:
\be
\delta M_{1/2} = \frac{m_{3/2}}{16\pi^2}\left(\frac{\Lambda^2}{M_P^2}+\dots\right)\,,
\ee
where $\Lambda$ is the cut-off scale and the dots represent subleading logarithmically divergent terms.
In string theory we expect that $\Lambda\leq M_s\sim M_P/\vo^{1/2}$ which then corrects the gaugino masses to order
$\delta M_{1/2}\leq \alpha_\lp M_P/\vo^2$ which is smaller than the sequestered gaugino masses $M_{1/2}\sim M_P/\vo^2$.

This behaviour of sequestered models motivated the work of Randall and Sundrum to introduce anomaly mediation.
However, as we will illustrate below, the approximate no-scale structure of LVS makes anomaly mediated corrections
to soft-terms subleading (they vanish identically for no-scale models) in generic points of parameter space.

\subsection{Anomaly mediated contributions}
\label{anosec}

In this appendix we examine anomaly mediated contributions to soft-terms and compute their strength
in the dS constructions discussed in Sec.~\ref{dsv}. The anomaly mediated gaugino masses~\cite{Bagger:1999rd} are given by\footnote{Note that there is a certain discussion on the validity of this formula~\cite{deAlwis:2008aq}. For the purpose of this paper we assume that the standard derivation from field theory or string theory~\cite{Conlon:2010qy} is valid.}
\be
M_{1/2}^{\rm anom}=\frac{g^2}{16 \pi^2}\left[\left(T_R-3T_G\right)m_{3/2}+\left(T_G-T_R\right)F^I\partial_I K
+\frac{2T_R}{d_R}F^I\partial_I\ln\det \tilde{K}_{\alpha\beta}\right],
\label{eq:anomalygauginomasses}
\ee
where $T_{G,R}$ are the Dynkin indices of the adjoint representation
and the matter representation $R$ of dimension $d_R$ (summation over all matter representations is understood).
Assuming that the K\"ahler metric for matter fields can be written as $\tilde{K}_{\alpha\beta}=\delta_{\alpha\beta} f_\alpha\tilde{K}$,
the expression (\ref{eq:anomalygauginomasses}) reduces to
\be
M_{1/2}^{\rm anom}=\frac{g^2}{16 \pi^2}\left[\left(T_R-3T_G\right)m_{3/2}+\left(T_G-T_R\right)F^I\partial_I K
+\frac{2T_R}{\tilde{K}} F^I\partial_I\tilde{K}+\frac{2T_R}{d_R}\sum_{\alpha=1}^{d_R} F^I\partial_I\ln f_\alpha\right].
\label{eq:anomalygauginomasses2}
\ee
In the local case, we find that the leading order anomaly mediated contribution can be written in terms
of the modulus dominated gaugino mass $M_{1/2}$ given in (\ref{ggmm})
\be
M_{1/2}^{\rm anom} = \frac{g^2}{16\pi^2} \left[\left(T_R - T_G\right)\left(1-s\beta^{U_i}\partial_{u_i} K_{\rm cs}\right)
- \frac{4 T_R}{\omega_S'} \left(c_s-\frac 13 \right)+\frac{2sT_R}{d_R}\sum_{\alpha=1}^{d_R} \partial_{s,u}\ln f_\alpha\right] M_{1/2},
\label{anorel}
\ee
with $\omega_S'$ as defined below (\ref{ds}). For the ultra-local case we find instead
\be
M_{1/2}^{\rm anom} = \frac{g^2}{16\pi^2} \left[\left(\frac{T_R}{3} - T_G\right)\left(1-s\beta^{U_i}\partial_{u_i} K_{\rm cs}\right)
+\frac{2sT_R}{d_R}\sum_{\alpha=1}^{d_R} \partial_{s,u}\ln h_\alpha\right] M_{1/2}\, .
\label{anorel2}
\ee
Therefore in both cases the anomaly mediated contribution is loop suppressed with respect to the moduli mediated one.
This result is the consequence of the approximate no-scale structure of LVS models.

A more careful analysis is needed for a very particular point in the underlying parameter space:
$\omega_S' \to 0$, i.e. in the very tuned situation where the F-term of the dilaton is vanishing at leading order because of a special compensation between the contribution to $F^S$ from $D_S W$ and $D_{T_b} W$. In this case the leading contribution to gaugino masses given in (\ref{ggmm}) is zero and the first non-vanishing moduli mediated contribution can be estimated to scale as $M_{1/2}^{\rm new}\sim m_{3/2} \sqrt{\ln\vo}/\vo $. On the other hand, the anomaly mediated contribution scales as $M_{1/2}^{\rm anom}\sim c \,M_{1/2}^{\rm new}$ where $c = c' \left(\frac{g^2}{16\pi^2}\right) \ln\vo$ and $c'$ denotes a numerical factor arising from evaluating~\eqref{anorel2} exactly.
For $g\simeq 0.1$ and $\vo \simeq 5\cdot 10^6$ (the value needed to get $M_{1/2}^{\rm new}$ approximately around the TeV-scale), $c$ is roughly of order $c'\times10^{-3}.$ Depending on the exact value of $c',$ which is beyond the scope of this analysis, we can achieve competing contributions from moduli mediation and anomaly mediation.

\subsection{Moduli redefinitions}

Desequestering can also potentially occur due to moduli redefinitions which might be necessary
order by order in perturbation theory. This desequestering effect can for example arise due to a shift of the local cycle $\tau_\SM\rightarrow \tau_\SM+\alpha\ln\vo$ which has the effect of making the soft-terms of the same order as the gravitino mass~\cite{Conlon:2010ji,Choi:2010gm}.

Such moduli redefinitions depend on the structure of the D-brane configuration. In particular, it has been argued that redefinitions are absent for configurations involving only D3-branes at orbifold singularities but are present for D3-branes at orientifold singularities
and in cases with both D3- and D7-branes (see~\cite{Conlon:2010ji}).

We emphasise that desequestering occurs only if the moduli redefinition leads to a change in the functional form of the K\"ahler potential. Arguments suggesting a change in the functional form were presented in \cite{Conlon:2010ji} but an explicit computation of such a change is still not available in the
literature. Some recent explicit computations of the K\"ahler potential \cite{Grimm:2013bha, Junghans:2014zla} have shown that perturbative corrections can be such that, along with a field redefinition, there is also an additional term generated in the K\"ahler potential. In these cases, however,
the two effects conspire to leave the functional form of the K\"ahler potential invariant.
More detailed studies of perturbative corrections to the K\"ahler potential are crucial to get a comprehensive understanding  of the relationship between moduli redefinitions and desequestering.

\subsection{Superpotential desequestering}

Apart from potentially destroying the hierarchy between soft masses and $m_{3/2}$,
various subleading effects can have important phenomenological consequences.
Interesting constraints arise from non-perturbative terms in the
superpotential involving visible sector fields \cite{Berg:2010ha}. Superpotential terms of the type
\be
\label{sud}
\hat{W} = \left( \hat{\mu} H_u H_d + \hat {\lambda}^u_{ij} Q^i u^j H_u
+ \hat {\lambda}^d_{ij}  Q^i d^j H_d + \hat {\lambda}^u_{ij} L^i e^j H_d \right) e^{-a_s T_s}\,,
\ee
would lead to flavour violation and CP-violation via A-terms with a strength sensitive to the hierarchy between soft masses and $m_{3/2}$.
For $M_{\rm soft}^2 \sim m_{3/2}^2/\vo^n$ the strength of CP and flavour violation induced by A-terms would be
\be
\label{flav}
\delta \sim  \vo^n   10^{-16} \left( {v \over 100 \phantom{.} {\rm GeV}} \right)\,,
\ee
with $v$ equal to the Higgs VEV. CP violation and FCNC bounds then require
$\vo < 10^5$. This gives a slight tension with our results but there can be several ways around this
issue. The estimate~(\ref{flav}) is based on effective field theory arguments; it assumes generic order one coefficients for the superpotential terms in (\ref{sud}).  A  string computation
of  the coefficients was done in \cite{Berg:2012aq}. This indicates that the coefficients are suppressed unless
the SM cycle and the cycle on which the instanton is supported share a homologous two-cycle. The presence of flavour symmetries \cite{1106.6039,1002.1790,1111.3047, 0810.5660} in the visible sector can also alleviate
this tension.

\bibliographystyle{JHEP}
\bibliography{biblio}
\end{document}